\documentclass[reprint,amsmath,amssymb,aps,nofootinbib]{revtex4-2}
\usepackage{graphicx}
\usepackage{float}
\usepackage{dcolumn}
\usepackage{xfrac}
\usepackage{bm}
\usepackage{tikz}
\usepackage{cancel}
\usepackage{ulem}
\usepackage[justification=justified,singlelinecheck=false]{caption}
\DeclareUnicodeCharacter{2212}{-}
\usetikzlibrary{shapes.geometric}
\usetikzlibrary{arrows.meta,arrows}
\usetikzlibrary{quotes,angles}
\usepackage{multirow}
\usepackage{caption}
\usepackage{ulem}
\usepackage{booktabs}

\begin{document}

\preprint{APS/123-QED}

\title{Searches for  {Extra} Higgs Bosons using $t\bar{t}+$Higgs{{$(\to b\bar b)$}} Events \\[0.15cm] within 2HDMs:
Direct versus Indirect Probes}

\author{Esteban Chalbaud$^{1}$}
\email{esteban.mogollon@lip.pt}

\author{Stefania De Curtis$^{2}$}
\email{decurtis@fi.infn.it}

\author{Luigi Delle Rose$^{3}$}
\email{luigi.dellerose@unical.it}

\author{Atri Dey$^{4}$}
\email{atri.dey@physics.uu.se}
\email{atridey1993@gmail.com}

\author{Stefano Moretti$^{4,5}$}
\email{stefano.moretti@cern.ch}

\author{António Onofre$^{1,6}$}
\email{antonio.onofre@cern.ch}

\affiliation{
{$^1$Laboratório de Instrumentação e Física Experimental de Partículas, Universidade do  Coimbra, 3004-516 Coimbra, Portugal}\\
{$^2$INFN – Sezione di Firenze \& Dipartimento di Fisica e Astronomia, Università di Firenze, Via G. Sansone 1, 50019 Sesto Fiorentino, Firenze, Italy}\\
{$^3$Dipartimento di Fisica, Universit\`a della Calabria, 87036 Arcavacata di Rende, Cosenza, Italy\\
INFN, Gruppo Collegato di Cosenza, Arcavacata di Rende, I-87036, Cosenza, Italy}\\
{$^4$Department of Physics and Astronomy, Uppsala University, Box 516, 751 20 Uppsala, Sweden}\\
{$^5$School of Physics and Astronomy, University of Southampton, Highfield, Southampton SO17 1BJ, United Kingdom}\\
{$^6$Centro de Física da Universidade do Minho e Universidade do Porto (CF-UM-UP), Universidade do Minho, 4710-057 Braga, Portugal}
}
\date{\today}

\begin{abstract}
\noindent
We study the possibility of establishing the production of   additional Higgs states in the process $gg,q\bar q\to t\bar t \Phi$, where $\Phi$ = $H_{2,3}$, with $H_2$ being CP-even and $H_3$ being CP-odd, at the Large Hadron Collider (LHC), by solely exploiting the kinematic features of the reconstructed $t\bar t$ system.  We adopt  as reference theoretical framework a generic  CP-Conserving  2-Higgs Doublet Model (2HDM), which also accommodate a Standard Model (SM)-like Higgs state $H_1$. We show that the masses $m_{H_{2,3}}$ exhibit clear correlations with the $t\bar{t}$ system properties and could, in principle, be extracted from these. Moreover, the CP properties of the $H_{2,3}$ states can  be determined, even when both states are produced simultaneously.   We then compare the results produced using this method with those obtained from a full kinematic reconstruction of the $H_{2,3}$ decays in the most studied $b\bar b$ channel (we take $m_{H_{2,3}}< 2m_t$),  thus proving the superiority of the  approach here proposed. This paves the way to both the discovery and characterization of additional Higgs states produced {\sl inclusively} in association with top-antitop quark pairs, thereby dispensing of the complications intrinsic to the   {\sl exclusive} reconstruction of such states from their  decay products. We test this by establishing the sensitivity of our approach in the case of a Composite 2HDM (C2HDM), describing  the Higges as pseudo-Nambu Goldstone Bosons (pNGBs) and naturally predicting Higgs mass spectra in the range of sensitivity of the described analysis.

\end{abstract}

\maketitle

\section{Introduction \label{sec:introduction}}

The top quark $t$ and the Higgs boson $h_{\rm SM}$ are the two  heaviest objects present in the Standard Model (SM). Their dynamics is strongly interlaced,  as evident from the so-called `hierarchy problem' of the SM, hence,  any modification of the Higgs sector, as described in Beyond the SM (BSM) scenarios, will alter their interaction, which is  governed by the Yukawa coupling ${y_t}_{ t\bar th_{\rm SM}}$. This can be modified also by the presence of extra fermions, like heavy top partners. Therefore, in a sense, the top and Higgs states of the SM may well act together as a portal to some underlying BSM physics.

There is no reason to believe that the discovered SM-like Higgs boson should be a unique state, when both matter - made up by quarks ($q$) as well as leptons ($\ell$) - and forces - mediated by Electro-Weak (EW) ($\gamma, W^\pm, Z$) and strong ($g$) gauge bosons - display various kinds of multiplicity (in flavor, color, charge, etc.). As the SM-like Higgs boson is a doublet and within the SM CP-Violating (CPV) effects are only of Cabibbo-Kobayashi-Maskawa (CKM) origin, it becomes natural to assume a CP-Conserving (CPC) 2HDM \cite{Branco:2011iw} as a possible construct hosting Higgs companion states. Specifically, herein, after EW Symmetry Breaking (EWSB), there would be five such states: three neutral ($H_{1,2,3}$) and two charged ($H^\pm$) ones. In particular, alongside one neutral Higgs state playing the role of the $h_{\rm SM}$, say, $H_1$ (which we take CP-even to comply with current experimental evidence), there are one additional such state, say, $H_2$, and also a CP-odd one, say, $H_3$. There would then be mixing between $H_1$ and $H_2$ altering the SM-like Higgs  interactions but there could also be the possibility of producing the $H_{2,3}$ states as real objects in the LHC 
detectors, as evidence of a 2HDM dynamics. While the aforementioned mixing is strongly constrained by LHC data involving the $H_1$ state,  
one of the most effective ways to produce its  companions, which might at once also enable one to establish their corresponding Yukawa couplings, i.e., the strength of the $t\bar tH_{2,3}$ interactions, would be the process $gg,q\bar q\to t\bar t \Phi$, where $\Phi=H_{2,3}$. 

While the case with $\Phi=H_1$ is under vigorous investigation at present by ATLAS and CMS \cite{ATLAS:2018urx,CMS:2020hwz, CMS:2018uxb}, the study of the cases with $\Phi=H_{2,3}$ has received considerable less attention (for obvious reasons, as such companion Higgs states are yet to be detected). We want to remedy here such a shortcoming. However, contrary to mainstream phenomenological approaches, which attempt to reconstruct the $H_{2,3}$ states {\sl directly} from their decays, we will address here the case in which their presence and properties (e.g.,  mass and CP quantum numbers) can potentially be inferred {\sl indirectly} from studying the system recoiling against these, i.e., the $t\bar t$ pair, along the lines of what done in Ref.~\cite{Azevedo:2022jnd}.

In fact, we will be able to prove the superiority of the latter approach
with respect to the former one in both enabling the detection of the $H_{2,3}$ states and establishing their properties discussed above. This is a powerful result, which can afford one with a better control of the underlying systematics, given that the $t\bar t$ system has extensively been studied at the LHC in many guises and the difficulties of searching for explicit decays of the companion Higgs states (in whatever final state, each with different selections and efficiencies, in presence of numerous backgrounds requiring dedicated modeling, etc.). We will therefore conclude that an indirect search of the kind above can be more effective than a direct one as we shall prove this to be true for one of the dominant decay $H_{2,3}$ channels, i.e., $b\bar  b$ (when $m_{H_{2,3}}<2m_t$, which we assume throughout). This therefore paves the way to test various realizations of the 2HDM, not just in terms of their Yukawa structures \cite{Branco:2011iw}, but also as Higgs sector representations of different BSM scenarios like Supersymmetry and Compositeness, thereby eventually contributing to separate the two hypotheses (as advocated in \cite{DeCurtis:2018iqd}). 

Our paper is organized as follows. After a description of our generic implementation of the 2HDM scenario, we illustrate the observables used to extract and characterize the $H_{2,3}$ states. This will be followed by a discussion of our analysis and corresponding results. We then summarize and conclude.

\section{Model implementation \label{sec:model}}

In this section we discuss the implementation of the 2HDM realizations that we tested.

\subsection{The Theory}
The scenarios that we considered are schematically characterized by the following structure:
\begin{equation}\label{reference}
{{\cal L}_\textrm{Extended}}={\cal L}_{\rm 2HDM} + {\cal L}_{d \ge 6},
\end{equation}
where ${\cal L}_{\rm 2HDM}$ has the same form as the Lagrangian of the 2HDM \cite{Branco:2011iw} and contains the kinetic terms,  scalar potential (up to quartic terms) and Yukawa interactions,
\begin{equation}
{\cal L}_{\rm 2HDM} = {\rm{kinetic~terms}} + V(\Phi_1,\Phi_2) +  {\cal L}_{\rm Yukawa}, 
\end{equation}
with $\Phi_1$ and $\Phi_2$ being the isospin scalar doublets and
\begin{align}\label{2HDM-potential}
V(\Phi_1, \Phi_2) & = m_1^2 \Phi_1^\dagger \Phi_1 + m_2^2 \Phi_2^\dagger \Phi_2 
- \left[m_3^2 \Phi_1^\dagger \Phi_2 + \text{h.c.} \right] \notag\\
& +\frac{\lambda_1}{2} (\Phi_1^\dagger \Phi_1)^2 + \frac{\lambda_2}{2} (\Phi_2^\dagger \Phi_2)^2 \notag\\
& + \lambda_3 (\Phi_1^\dagger \Phi_1)(\Phi_2^\dagger \Phi_2) 
+ \lambda_4 (\Phi_1^\dagger \Phi_2)(\Phi_2^\dagger \Phi_1) \notag\\
& + \frac{\lambda_5}{2} (\Phi_1^\dagger \Phi_2)^2 
+ \lambda_6 (\Phi_1^\dagger \Phi_1)(\Phi_1^\dagger \Phi_2) \notag\\
& + \lambda_7(\Phi_2^\dagger \Phi_2)(\Phi_1^\dagger \Phi_2) + \text{h.c.} 
\end{align}
being the Higgs potential. As intimated, notice that we take the $\lambda$'s and masses to be real, so that our underlying BSM scenario is CP-conserving and we will    identify the emerging (neutral) Higgs mass eigenstates $H_{1,2}$ as CP-even (with $m_{H_1}<m_{H_2}$) and $H_3$ as CP-odd (we also have two charged Higgs states, $H^\pm$).
The $ {\cal L}_{d \ge 6}$ part  includes  operators (starting from dimension 6) that can generate modifications to the Higgs couplings to bosons and fermions, hence, consequent effects in specific experimental observables in Higgs physics as well as global EW precision tests. In general, these  operators generate corrections that are suppressed by some high energy scale, $\Lambda$, via terms of ${\cal O}(v^2/\Lambda^2)$, with $v$ being an EW scale parameter connected to the Higgs doublet Vacuum Expectation Values (VEVs).

The importance of targeting 2HDMs for our purposes is twofold. On the one hand, they are the minimal setup allowing for the presence of both a CP-even and CP-odd Higgs state (in addition to the SM-like one). On the other hand, they can be naturally realized in fundamental theories of the EW scale, such as, e.g., Supersymmetry, which embeds fundamental Higgs bosons, or in Composite Higgs Models (CHMs), which adopt pseudo-Nambu Goldstone Bosons (pNGBs). Concrete
realizations of these scenarios based on a 2HDM are the Minimal Supersymmetric Standard Model (MSSM), see Ref.~\cite{Moretti:2019ulc} for a review, and the  
Composite 2HDM (C2HDM), as given in Ref.~\cite{DeCurtis:2018zvh}, respectively.  In fact, these   two constructs can predict different deviations from the SM Higgs sector, potentially enabling one to separate them by using collider observables: while Ref.~\cite{DeCurtis:2018iqd} concentrated on differences emerging in gauge couplings of 2HDM states, here, we concentrate on those stemming from the Yukawa sector.

The 2HDM extends the Higgs sector but leave the hierarchy problem untouched. All (pseudo)scalar Higgs states are treated as fundamental, so quantum corrections still drive the Higgs mass towards very high scales (the discussed hierarchy problem).  Supersymmetry solves this problem by including the contributions of the SUSY partners. Another elegant remedy, given by the Compositeness hypothesis,  is to assume that the SM-like Higgs and its 2HDM partners are not elementary. Instead, they are composite fermion-antifermion bound states of a new strong force, much like pions in QCD.  By embedding the composite sector in a larger global symmetry that spontaneously breaks at the (compositeness) scale 
$f\sim$ TeV, the Higgs doublets emerge as pNGBs.
The pNGB nature protects their masses from large radiative corrections and 
explicit symmetry breaking interactions generate a calculable Higgs potential via loops. The lightest Higgs mass can naturally be set at 125 GeV with values of 
$f$  in the TeV range
while  new resonances (spin-0, -1/2 and -1) can appear near $f$, offering LHC search targets. Finally, Higgs couplings deviate subtly from the SM ones, providing room to satisfy EW precision tests depending upon the actual companion Higgs masses.

We will not deal with Supersymmetry here, as we will refer to the viable construction of the C2HDM performed in Ref.~\cite{DeCurtis:2018zvh} (see also Refs.~\cite{DeCurtis:2019jwg,DeCurtis:2023ucs}). The effective Lagrangian describing the interactions of the pNGBs of the SO(6)/SO(4)$\times$SO(2) coset with the SM particles and the composite extra fermions is expressed in terms of the strong sector  parameters. We performed a scan over these by requiring to reconstruct the measured values for the EW VEV, the Higgs mass and the top mass.  We then subjected the model parameter space  to the most up-to-date  theoretical and experimental constraints.
For the present analysis we are interested in values of the additional neutral Higgses in the range $m_{H_1}<m_{H_{2,3}}<2~ m_t$. Finally, among them,  we selected the ones with the largest Yukawa couplings $y_{\bar t t H_{2,3}} $, to maximize the number of signal events. Furthermore, since we want to compare the present method with the one exploiting a full
kinematic reconstruction of the $H_{2,3}$ decays in the most studied $b \bar b$ channel, we selected C2HDM points with sizable BR($H_{2,3}\to \bar b b$). Three of them are presented in Table~\ref{table1}.

\begin{table}[h]
    \centering
    \begin{tabular}{|p{0.6cm}|p{1.0cm}|p{1.7cm}|p{2.0cm}|p{2.1cm}|}
        \hline  

         &    $f$ (GeV) &$m_{H_2}, m_{H_3}$ (GeV) & $y_{t\bar tH_2}/y_{t\bar th_{\rm SM}}$, $y_{t\bar tH_3}/y_{t\bar th_{\rm SM}}$ & BR($H_2\to b\bar b$), BR($H_3\to b\bar b$)\\
        \hline
    \hline
        BP1 & $~$ 860 & 286.2, 276.0 & $~$0.278, $~$0.348 & $~$ 0.432, 0.517 \\
        \hline
        BP2 & $~$ 850 & 307.0, 271.0 & $~$0.374, $~$0.371 & $~$ 0.340, 0.537 \\
        \hline
        BP3 & $~$ 920 & 322.0, 271.7 & $~$0.352, $~$0.369 & $~$ 0.169, 0.542 \\
        \hline
    \end{tabular}
    \caption{Benchmark Points (BPs) of the C2HDM. The  parameters relevant for the analysis are indicated, namely, the $H_{2,3 }$ masses, Yukawa couplings and Branching Ratios (BRs) in $b \bar b$.}
    \label{table1}
\end{table}

\subsection{The Tools}

We use~\texttt{MadGraph5\_aMC@NLO}~\cite{Alwall:2014hca} to generate signal events from the associated production of pairs of top quarks with both CP-even and CP-odd Higgs states ($p p \rightarrow t \bar{t} H_{2,3}$), with a private implementation of the~\texttt{2HDM} UFO model~\cite{Degrande:2014vpa} with no Flavor Changing Neutral Currents (FCNCs). We fixed the masses of the CP-even and CP-odd Higgs bosons to $m_{H_{2}}=180$~GeV and $m_{H_{3}}=150$~GeV,  respectively, as our reference for implementing the analysis strategy. These mass values are among the lowest ones from the C2HDM scan discussed above and should, in principle, be the most kinematically favored at the LHC. 
The corresponding Yukawa couplings normalized to the SM value were fixed to $-0.254$ (for the $H_{2}$) and $-0.197$ (for the $H_{3}$). However, notice that such a reference point is merely used as a seed of our kinematical analysis, as further explained below.

For completeness, we also generated with~\texttt{MadGraph5\_aMC@NLO} the most significant SM backgrounds, i.e., the $t \bar{t}$, single-top ($tb$, $tq$ and $tW^\pm$ channels), $t\bar{t}+V$ ($V$=$W^\pm$,$Z$), $t\bar{t}h_{\rm SM}$, $W^\pm/Z$+jets (including $b\bar{b}$) and diboson $(W^+ W^-, Z Z, W^\pm Z)$ processes. All Leading Order (LO) signals and Next-to-LO (NLO) backgrounds were generated for a center-of-mass energy $\sqrt{s}= 13.6$~TeV for the LHC.

We set the top quark and $W^\pm$ boson masses to $m_t = 172.5$~GeV and $m_W = 80.4$~GeV, respectively. Further, we utilize the \textbf{\texttt{NNPDF2.3}} Parton Distribution Functions (PDFs)~\cite{Ball:2012cx}. The renormalization and factorization scales were both dynamically defined as the sum of the transverse momenta of all final state particles, following the approach of Ref.~\cite{Azevedo_2020}.

{{Particle decays, including spin correlation effects for signal and background, were managed by {\texttt{MadSpin}}~\cite{Artoisenet:2012st}, assuming fully leptonic top quark decays in the signal. \textbf{\texttt{Pythia}}~\cite{Sjostrand:2006za} was used for parton showering, hadronization and heavy flavor decays. Starting from LO, in the case of $H_{2,3}$, or NLO, in the case of $h_{\rm SM}$, events at the parton level, we matched the emission from the latter to radiative matrix elements by using the MLM merging scheme~\cite{Alwall:2014hca}. A fast detector simulation was performed with {\texttt{Delphes}~\cite{deFavereau:2013fsa} using the default ATLAS configuration. Jets were clustered with the anti-$k_t$ algorithm ($\Delta R=0.4$) via \textbf{\texttt{FastJet}}~\cite{Cacciari:2011ma}. At the generation level, we required all partons and leptons to have $p_T > 10$~GeV with the pseudorapidity constraint $|\eta| < 5.0$ for both partons and leptons}.}}

The final analysis was conducted within the \texttt{MadAnalysis5} framework~\cite{Conte:2012fm}. Event selection required jets and leptons with $p_T > 20$~GeV and $|\eta| < 2.5$.  To reconstruct the $t\bar{t}$ system, a Multi-Variate Analysis (MVA) was employed to correctly assign jets to their parent quarks. Consistently with previous work~\cite{Azevedo_2022,Azevedo:2023xuc,Chalbaud:2024jsr}, Boosted Decision Trees (BDTs) were found to be the most effective classifiers (thus are being used in our forthcoming  analysis).

{{A kinematic fit, based on the method of Ref.~\cite{Azevedo_2020}, was used to reconstruct the di-lepton system in $t \bar{t} \rightarrow b \ell^{+} \nu_{\ell} \bar{b} \ell^{-} \bar{\nu}_{\ell}$ events, where $\ell=e,\mu$. This fit imposed mass constraints on the $W^\pm$ boson and top (anti)quark decay products while assuming that the total missing transverse energy ($\cancel{E_T}$) originates entirely from neutrinos.}}

\section{The $t\bar{t}$ System \label{sec:ttbar}}

In this section we describe how to reconstruct the decayed $t$ and $\bar t$ states in the presence of an additional $X$ object in the hard collision, in turn decaying into $ b\bar b$ only  (i.e., with 100\% BR)~\footnote{At the interpretation stage within the C2HDM, this assumption will of course be relinquished.}. Finally, we will introduce some observables making use of the corresponding ${\vec p}_t$ and ${\vec p}_{\bar t}$ three-momenta useful to extract the CP properties of the $X$ state.

In events where an additional boson \( X \) is radiated off a top-antitop pair, the reconstruction of the \( t\bar{t} \) system remains a central task for studying the underlying dynamics. To clearly understand the performance of the kinematic reconstruction, the procedure follows closely that of the \( t\bar{t}H_{2,3} \) case described in detail in Section~\ref{sec:generation}.B, where jets and leptons are matched to their parent partons and the kinematic configuration is fully reconstructed under known mass and momentum constraints. While the additional boson \( X \) is not explicitly reconstructed, its presence impacts the top-antitop pair kinematics and opens the door to the use of CP and mass sensitive observables.

\vspace{2mm}

In particular, we start by considering two angular observables, \( b_2 \) and \( b_4 \), defined in terms of the top and antitop three-momenta in the laboratory (LAB) frame:
\begin{equation}
\begin{aligned}
& b_2=\frac{(\vec{p}_t \times \hat{k}_z) \cdot(\vec{p}_{\bar{t}} \times \hat{k}_z)}{\left(\left|\vec{p}_t\right| \cdot\left|\vec{p}_{\bar{t}}\right|\right)}, \\
& b_4=\frac{p_t^z \cdot p_{\bar{t}}^z }{\left(\left|\vec{p}_t\right| \cdot\left|\vec{p}_{\bar{t}}\right|\right)},
\end{aligned}
\end{equation}
where \( \hat{k}_z \) is the beam direction unit vector. These observables are sensitive to the spin correlations and relative orientation of the top (anti)quarks and are thus affected by the CP properties of the \( X \) particle. Their distributions can show striking differences depending on whether \( X \) is a CP-even (scalar) or a CP-odd (pseudoscalar) state, even without direct access to \( X \)'s decay products. 

\begin{figure*}
\begin{center}
\begin{tabular}{ccc}
\hspace*{-5mm}\includegraphics[height=8.0cm]{ 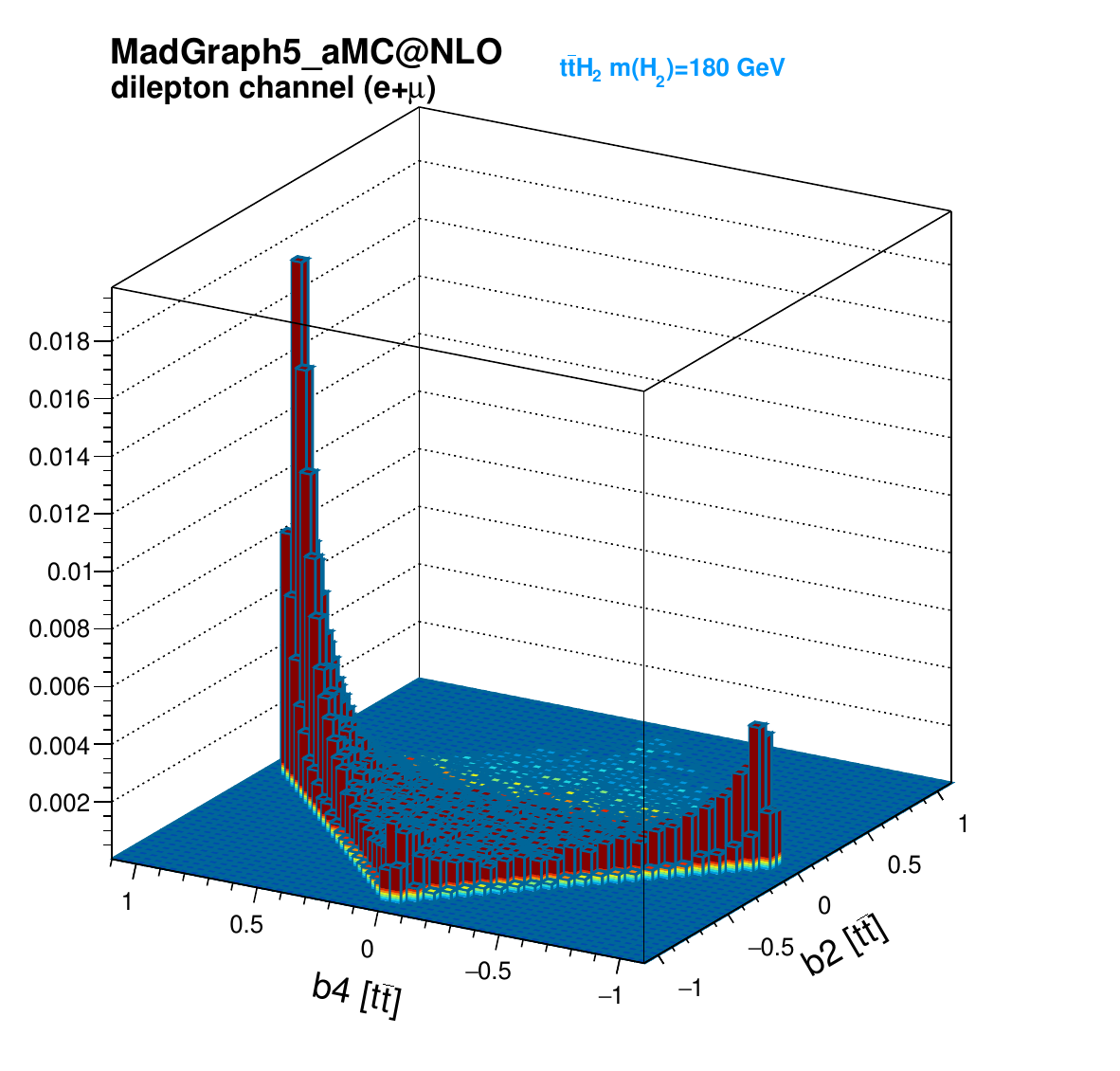}
\hspace*{2mm}\includegraphics[height=8.0cm]{ 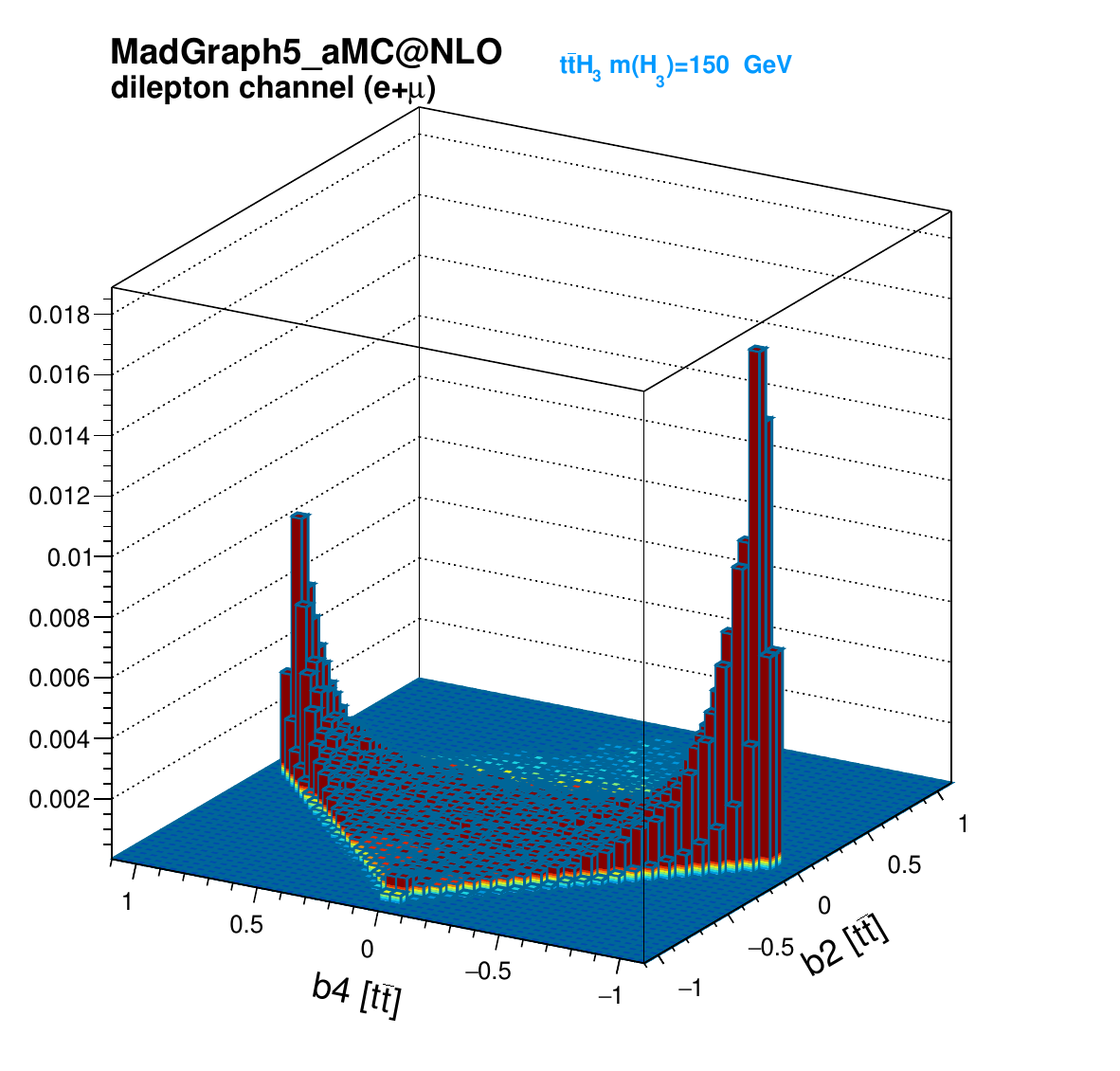}\\
\hspace*{-5mm}\includegraphics[height=8.0cm]{ 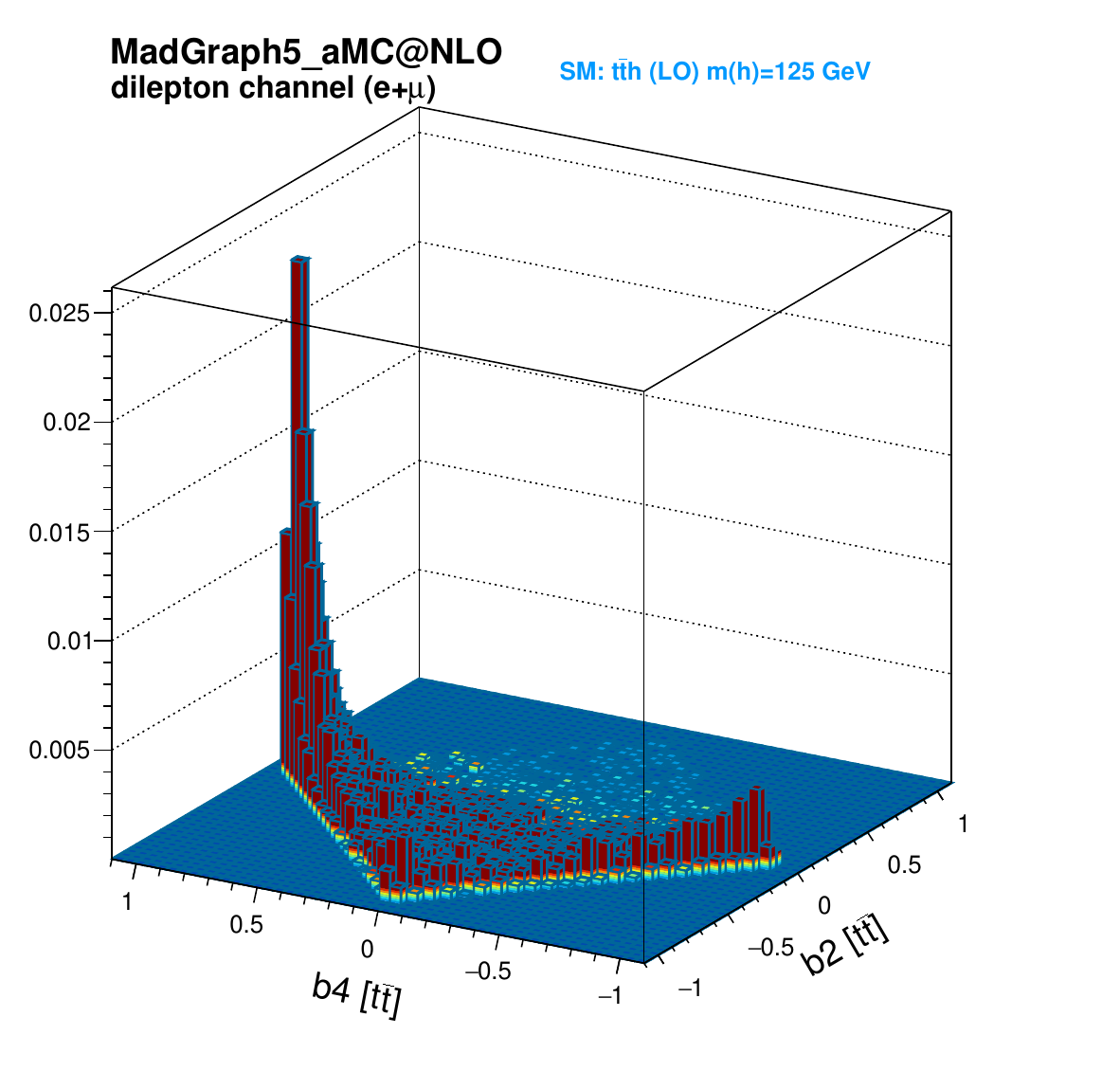}
\hspace*{2mm}\includegraphics[height=8.0cm]{ 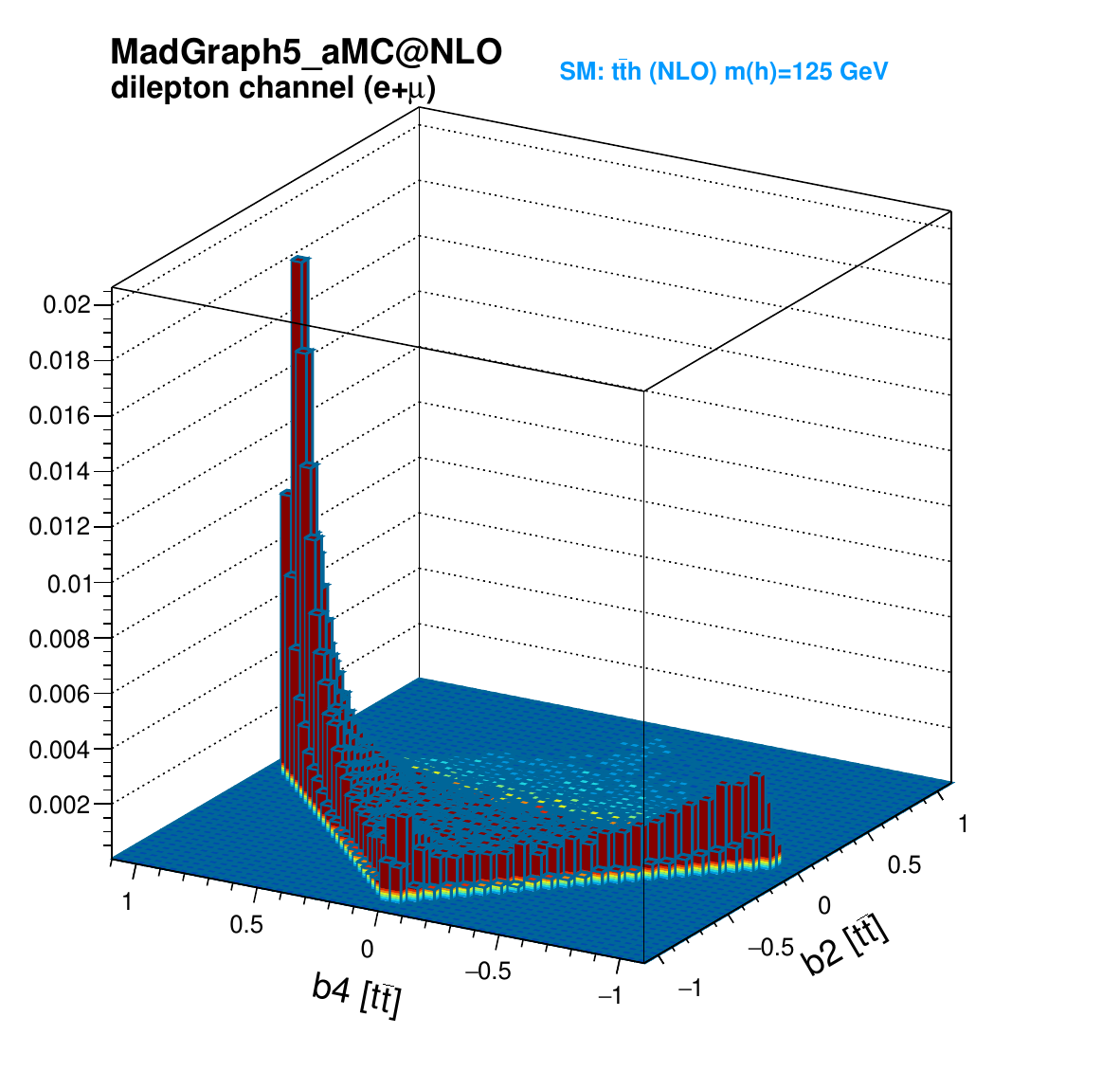}\\
\end{tabular}
\end{center}
\caption{Distributions of $t\bar{t}H_2$ (top-left) and $t\bar{t}H_3$ (top-right) for $b_2$ and $b_4$ are shown in the ($b_2$,$b_4$) plane, evaluated in the LAB frame at parton level at LO. In the bottom-left(right) plot we show $t\bar{t}h_{\rm SM}$ case at parton level at LO(NLO).}
\label{fig:b2_vs_b4}
\end{figure*}

\begin{figure*}
\begin{center}
\begin{tabular}{ccc}
\hspace*{-5mm}\includegraphics[height=8.0cm]{ 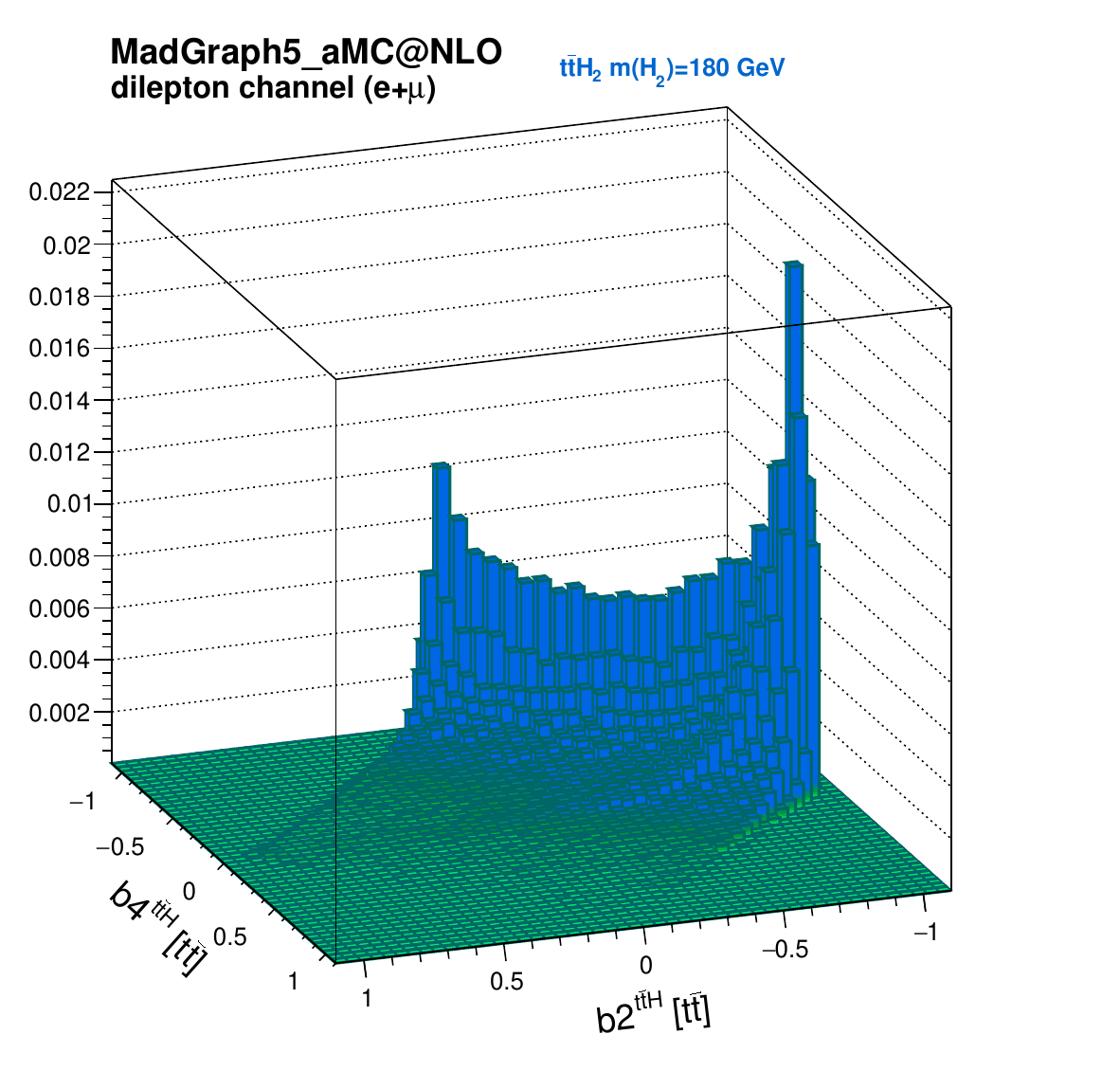}
\hspace*{2mm}\includegraphics[height=8.0cm]{ 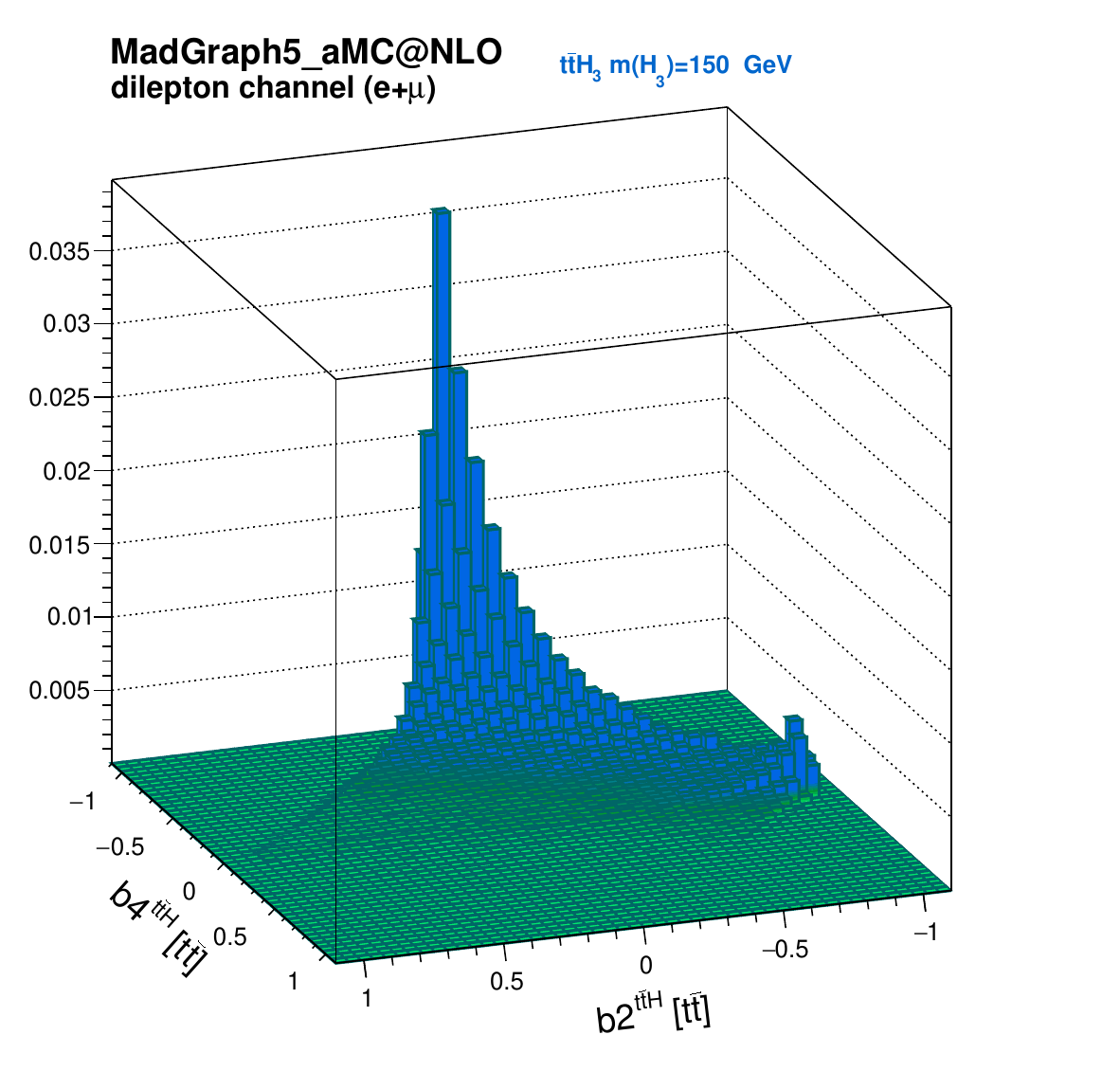}\\
\hspace*{-5mm}\includegraphics[height=8.0cm]{ 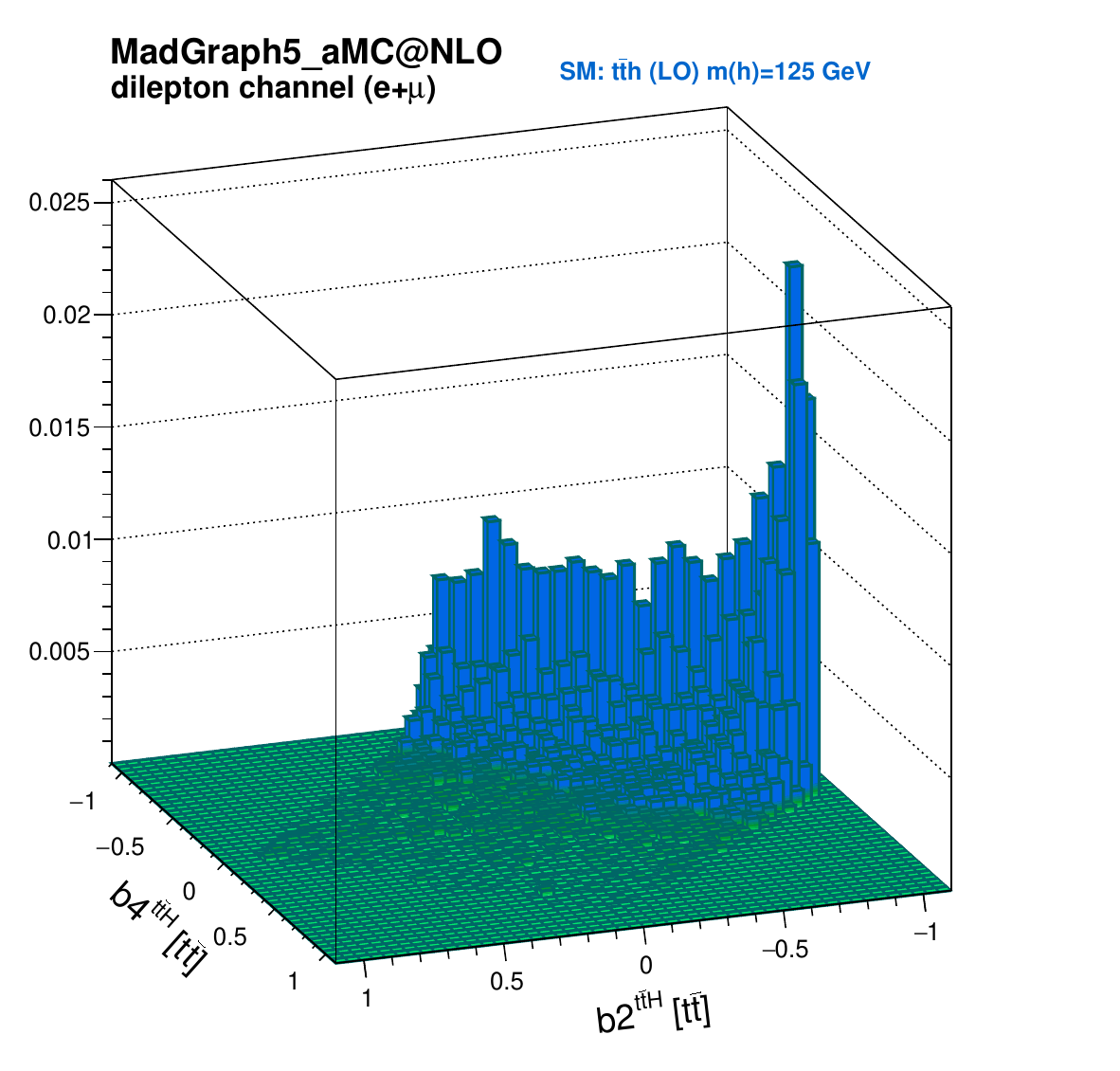}
\hspace*{2mm}\includegraphics[height=8.0cm]{ 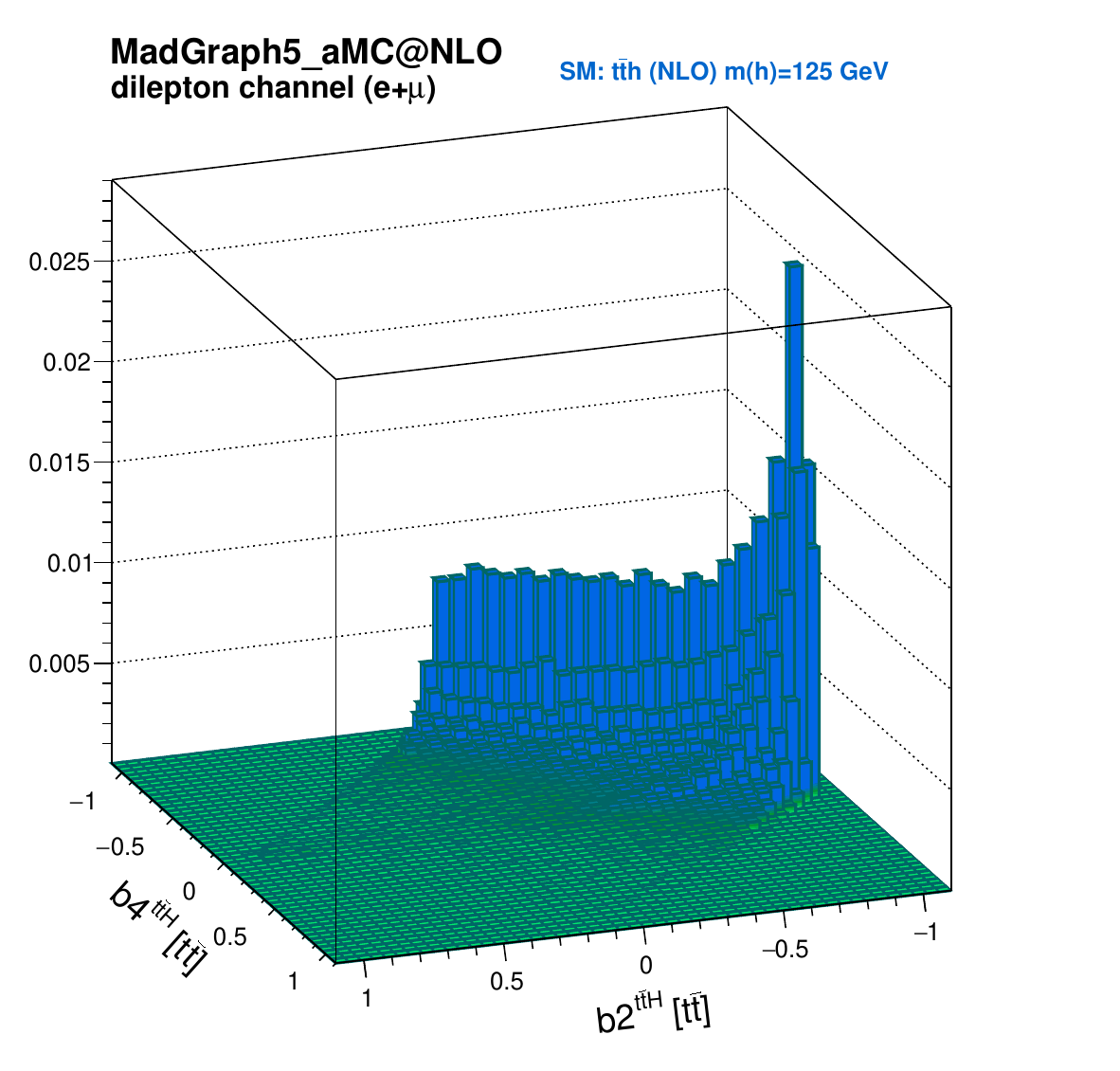}\\
\end{tabular}
\end{center}
\caption{Same as as in Figure~\ref{fig:b2_vs_b4} but using the CoM frame. }
\label{fig:b2_vs_b4_ttH}
\end{figure*}

This is shown in Figure~\ref{fig:b2_vs_b4} (top two plots), wherein the 2D distributions of $b_2$ against $b_4$ at parton level at LO, in the LAB frame, clearly make the point that these are significantly different in shape between the CP-even ($H_2$) and CP-odd ($H_3$) case, so that not only the individual shapes should be distinguishable (i.e., when $H_2$ or $H_3$ is detected on its own) but also their superposition should be traceable  (i.e., when $H_2$ and $H_3$ are detected simultaneously).  A similar result is obtained in the Center-of-Mass (CoM) frame, shown in Figure~\ref{fig:b2_vs_b4_ttH} (top two plots). here, it is worth mentioning that the distributions are here rotated around the $z$-axis (vertical axis) when compared with the previous ones.

Notice that in both figures  the distributions shown correspond to the following combination of masses: $m_{h_2}=180$ GeV and
$m_{h_3}=150$ GeV. We have verified that the mass difference between the $H_2$ and $H_3$ states is not responsible for the effects seen, namely $b_2$ and $b_4$ are almost independent of the mass of the $X$-particle produced in association with the $t \bar t$ pair.
As mentioned previously, this choice of masses was guided by the C2HDM scan performed using the most recent constraints, by selecting among the lowest allowed values for  $m_{H_{2,3}}$. Such a choice provides a good reference point for the analysis, since the $H_{2,3}$ production at the LHC may be kinematically favored and the variables used are largely independent of specific mass choices. It is also worth mentioning that values of the cross sections we take for the two processes $gg\to t\bar tH_{2,3}$ are not relevant either, once our results are presented in a model independent approach in terms of number of events, in turn function of the (squared) Yukawa couplings. We will come back to this later on in our analysis when discussing detector level results. For the forthcoming parton level ones, their normalization can be taken as arbitrary.  
 The bottom two plots of these figures show the corresponding distributions for the case of $t\bar t h_{\rm SM}$ production, again at parton level but now both at LO and NLO, from where it is indeed clear (by comparing $H_2$ and $h_{\rm SM}$ data at LO, i.e., for the CP-even case) that the mass plays no significant role~\footnote{Albeit not shown here, the same happens for the CP-odd case.}. Furthermore, by comparing the bottom two plots, it is  evident that NLO effects are also minimal. 
 
 {{The scalar ($H_2$) and pseudoscalar ($H_3$) nature of the Higgs bosons produced in association with the $t\bar{t}$ system  show clear and distinct effects in the $t\bar{t}$ kinematics. This makes the $b_2$ and $b_4$ variables, evaluated in the LAB or in the CoM reference frames, excellent observables to probe the CP nature of the recoiling Higgs bosons, even with masses that are not too far apart, as is the case here. 
}} 
In short, we are equipped with two powerful observables defined within the $t\bar t$ system yet sensitive to the CP properties of the recoiling one, irrespectively of its mass.

\begin{figure*}
\begin{center}
\begin{tabular}{ccc}
\hspace*{-5mm}\includegraphics[height=6cm]{ 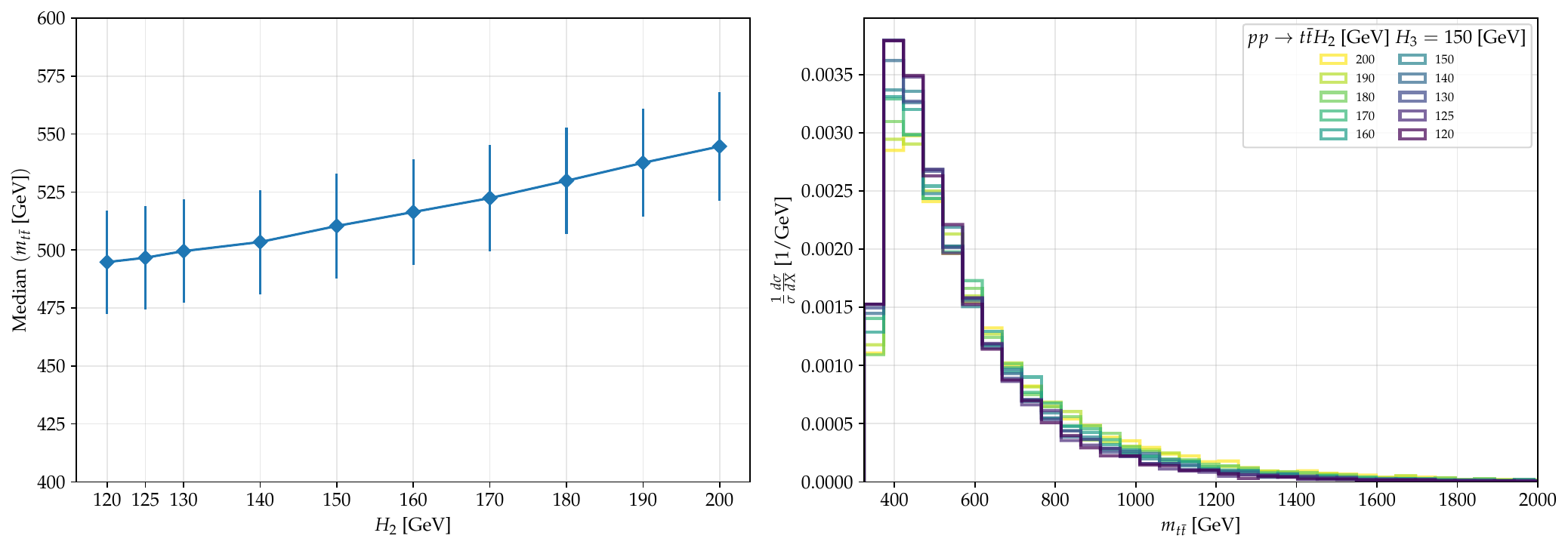}\\
\hspace*{-5mm}\includegraphics[height=6cm]{ 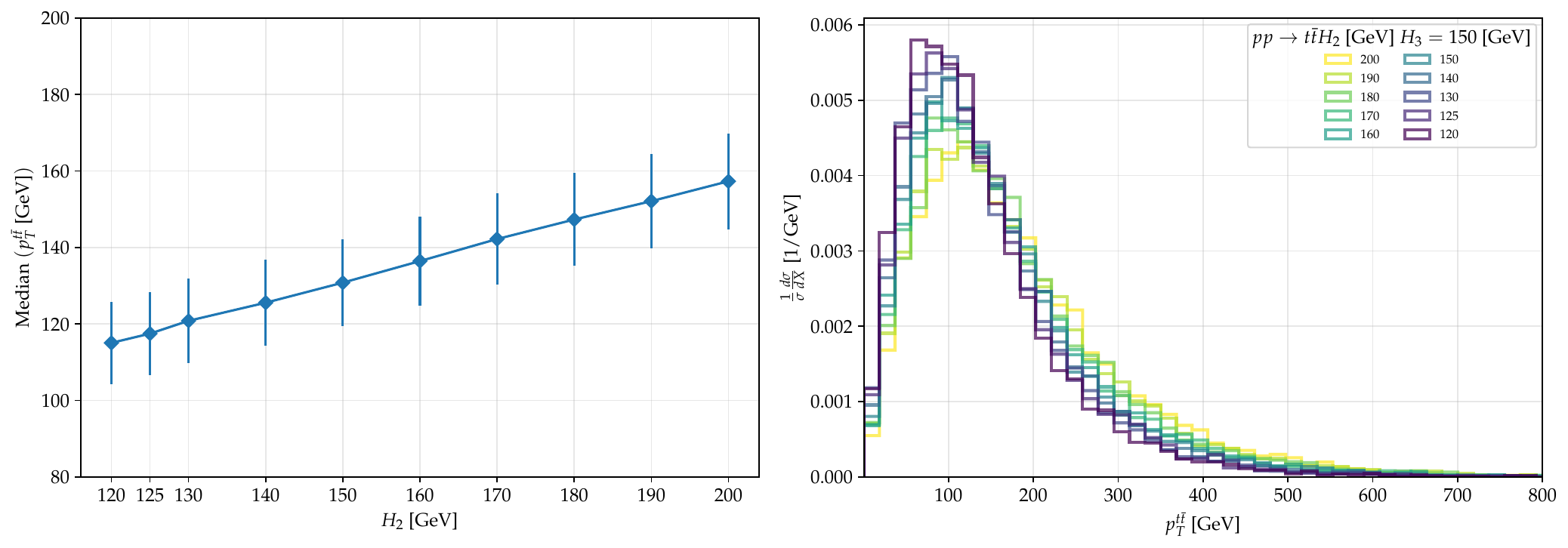}\\
\end{tabular}
\end{center}
\caption{Sensitivity of key kinematic observables to the mass of the additional CP-even Higgs boson ($m_{H_2}$) in $pp \to t\bar{t}H_2$ events, fixing the mass of the $H_3$ to $m_{H_3}=150$~GeV. (Left) The median (50th) of the $m_{t\bar{t}}$ (top) and $p_T^{t\bar{t}}$ (bottom) distributions are shown as a function of $m_{H_2}$. The uncertainty bars represent the precision with which the medians are extracted from the current simulation (see the text for details). (Right) The corresponding distributions normalized to the cross section for several $m_{H_2}$ values, illustrating the sensitivity of the observables to  $|m_{H_2}-m_{H_3}|$.}
\label{fig:mttbar_ptttbar_quartiles}
\end{figure*}

As a next step, we introduce two variables pertaining to the (anti)top quark pair that are instead sensitive to the mass of the recoiling object. These are the transverse momentum and invariant mass of the $t\bar t$ system:
\begin{equation}
\begin{aligned}
& p_T^{t\bar t}=\sqrt {(p^x_t+p^x_{\bar t})^2+
                      (p^y_t+p^y_{\bar t})^2}, \\
& m_{t\bar t}=\sqrt {(E_t+E_{\bar t})^2- (\vec{p}_t+\vec{p}_{\bar t})^2},
\end{aligned}
\end{equation}
which are frame invariant. In Figure~\ref{fig:mttbar_ptttbar_quartiles} we show the $m_{t\bar t}$ (top-right) and $p_T^{t\bar t}$ (bottom-right) distributions for different $H_2$ and $H_3$ mass gaps. We consider  the usual $H_3$ mass equal to 150~GeV, but, to make our point clear, we have now allow the $H_2$ mass to range from 150 to 250~GeV (in steps of 20~GeV), to check the dependence of these observables with the recoiling mass system. 
As shown in Figure~\ref{fig:mttbar_ptttbar_quartiles}, both observables are clearly affected by the mass of the recoiling object, even for small mass gaps. 
We quantify this effect by examining how the medians (50th percentile) of the distributions evolves with the recoil mass. These are also shown in Figure~\ref{fig:mttbar_ptttbar_quartiles}, for $m_{t\bar t}$ (top-left) and $p_T^{t\bar t}$ (bottom-left). The uncertainty bars indicate the precision with which the median can be extracted from the currently available simulation. Obviously, larger statistics would improve the determination of the correlations (i.e., the slopes of the median curves), however, a clear trend is already visible. This strong connection between the properties of the recoiling system and those of the $t\bar{t}$ system should be further explored in future LHC studies.
We have thus found two further powerful observables defined within the $t\bar t$ system yet sensitive to properties of the recoiling one, the Higgs masses in this case. 

We can now proceed with the description of $t\bar tH_{2,3}$ events.

\section{Associated $t\bar{t}$+Higgs Production \label{sec:generation}}

In this section we look into $t\bar{t}H_{2,3}$ production at the LHC, still focusing on the $t\bar{t}$ system and its properties. The goal here is to exploit the results obtained in the previous section, but allowing for both particles ($H_{2,3}$) to be produced simultaneously, i.e., checking how much their presence affects the kinematic properties of the $t\bar{t}$ system. We start by showing some relevant properties of the events, their selection as well as reconstruction (few example, of angular distributions) and finally discuss the results obtained.

\subsection{Properties of $t\bar{t}H_{2,3}$ Events}

The kinematic properties of separate $t\bar{t}H_2$ and $t\bar{t}H_3$ events are quite unique and distinct when looked from the perspective of the $t\bar{t}$ system alone. Indeed, the presence of a scalar or a pseudoscalar boson with distinct CP-properties do affect the $t\bar{t}$ kinematics, their transverse and longitudinal momenta, the phase space of the top quarks themselves and observables related to them. However, the parton level results from the previous session need to be validated by a proper Monte Carlo (MC) analysis at detector level.

\subsection{Event Selection and Reconstruction}

Following event generation, parton shower, hadronization and heavy flavor decays, $t\bar{t}H_{2,3}$ and (SM only) background events were passed through {\tt Delphes} to simulate a parametrized response of a typical LHC detector. 
Events were then pre-selected by requiring final states with at least four jets ($N_{\rm j} \ge$~4) and two opposite-charge leptons ($N_{\ell} \ge$~2, $e$ or $\mu$), as mentioned, with transverse momenta ($p_T$) above 20~GeV and pseudorapidities ($\eta$) below 2.5 in modulus (see Table~\ref{tab:cutflow}, Pre-Sel. column). The pre-selection efficiencies obtained are 18.1\% for $t\bar{t}H_{2}$ and 18.6\% for $t\bar{t}H_{3}$. 
In order to further reconstruct the events, we firstly attempt to match the jets to the corresponding partons from the heavy particles' decays, followed by the full reconstruction using several kinematical constraints.

\vspace*{2mm}
{\it Matching Jets to Partons}
\vspace*{2mm}

\begin{figure*}
\begin{center}
\begin{tabular}{ccc}
\hspace*{-5mm}\includegraphics[height=9.5cm]{ 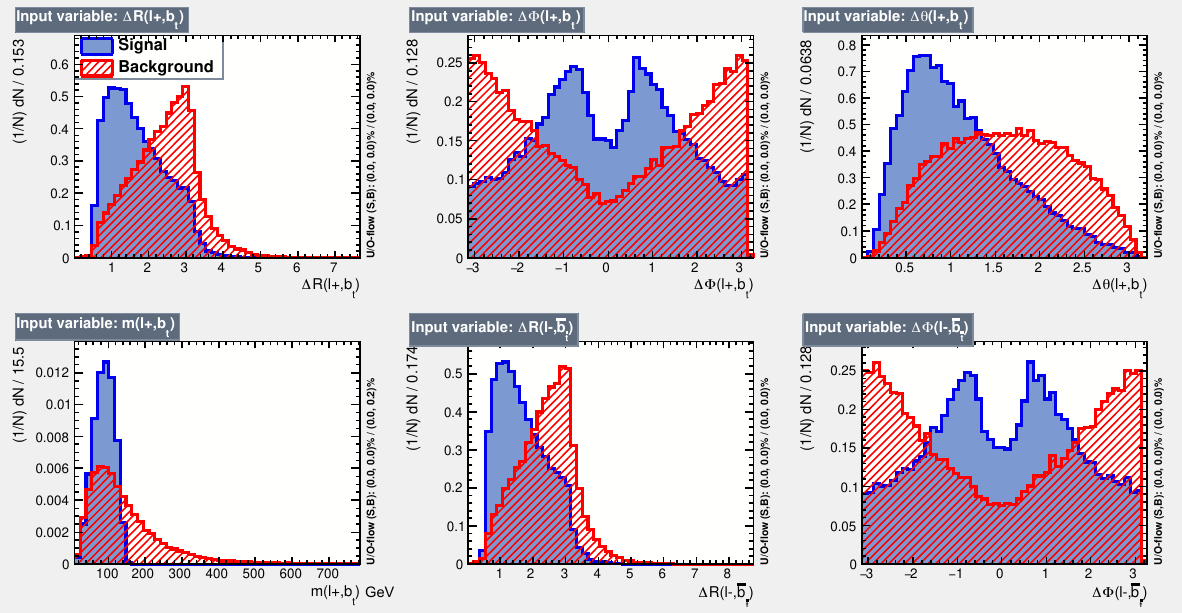}\\
\end{tabular}
\end{center}
\caption{Examples of distributions between partons and charged leptons. The distributions where particles are coming from the same (wrong) mother particle, are represented in Blue (Red) and labeled Signal (Background). Distributions are normalized with their bin size.}
\label{fig:TMVA_variables}
\end{figure*}

We now reconstruct the signal from $t\bar{t}H_{2,3}$ events, with $H_{2,3}\to b\bar b$  and $t\bar t $ decaying into di-leptonic final states by attempting to correctly match jets to the corresponding $b$-quark partons (originated in the decays of the $t$, $\bar{t}$ and the $H_{2,3}$ bosons), using the ROOT Toolkit for MVA (TMVA)~\cite{hoecker2009tmvatoolkitmultivariate}. We use angular distributions between the partons and charged leptons as input variables to the TMVA. The $\Delta R$, $\Delta\phi$, $\Delta\theta$ and invariant mass distributions of the pairs ($\ell^+$,$b_t$), ($\ell^-$,$\bar{b}_{\bar{t}}$) and ($b_{H_{2,3}}$,$\bar{b}_{H_{2,3}}$) are probed to check how likely is that a specific combination of jets and leptons correctly matches parton level kinematics. In Figure~\ref{fig:TMVA_variables}, we show a few examples of these distributions. The ones where particles are coming from the same (wrong) mother particle, are represented in blue (red) and labeled Signal (Background). 

\begin{figure*}
\begin{center}
\begin{tabular}{ccc}
\hspace*{-5mm}\includegraphics[height=6.5cm]{ 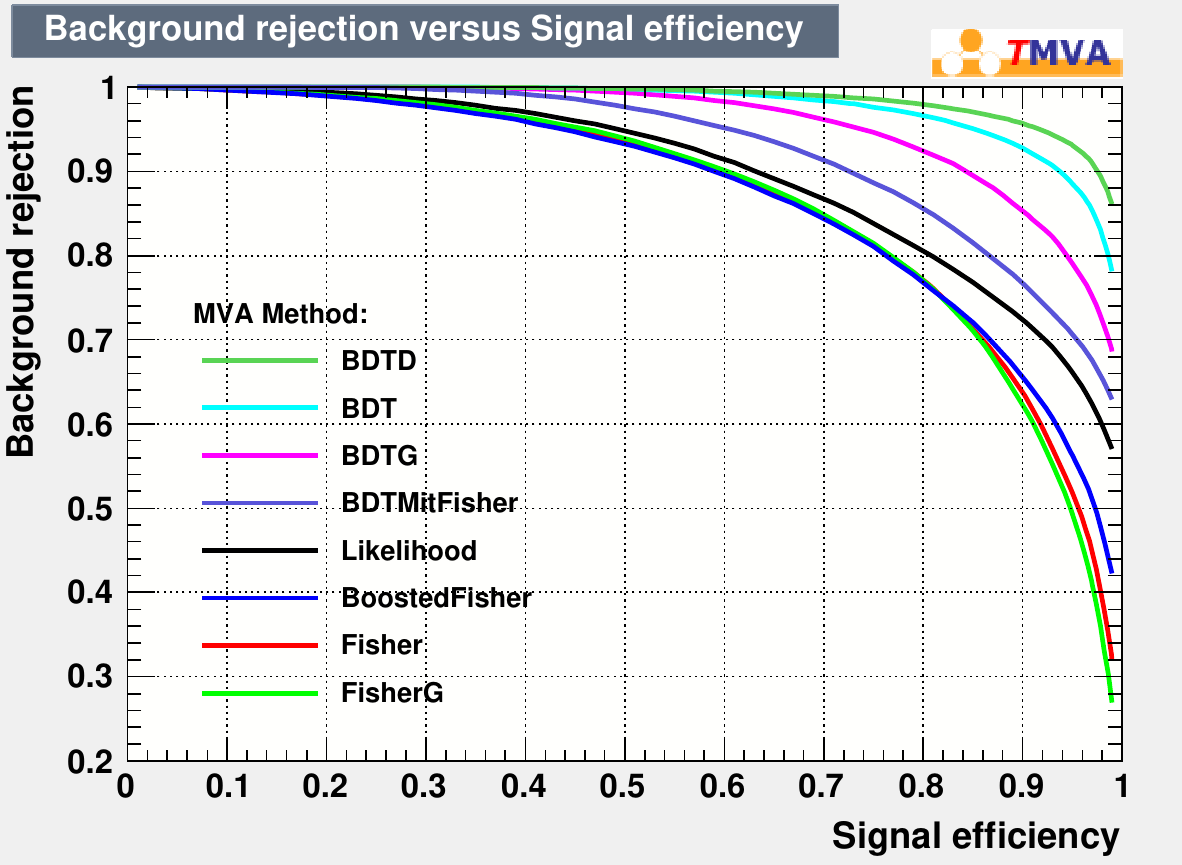}
\hspace*{5mm}\includegraphics[height=6.5cm]{ 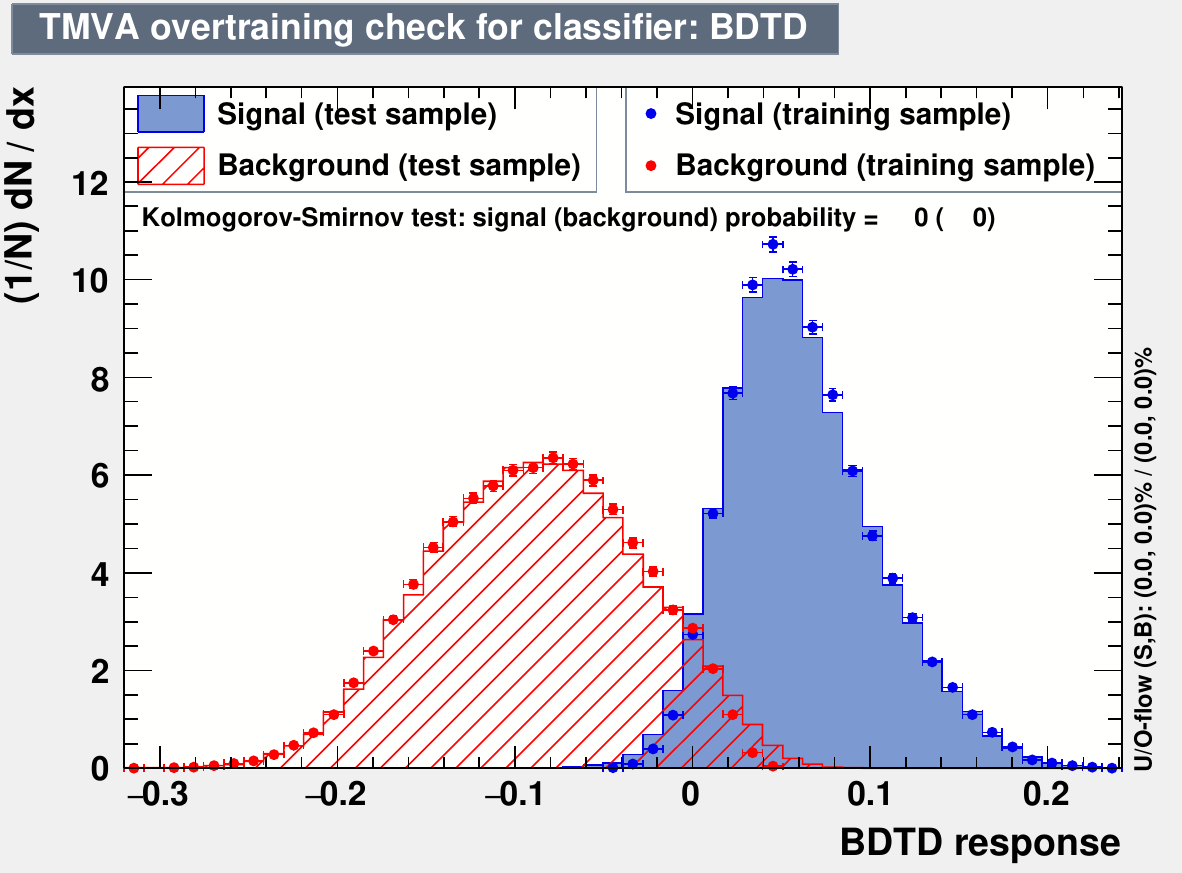}\\
\end{tabular}
\end{center}
\caption{The Receiver Operating Characteristic (ROC) curve for several multivariate algorithms (left) and the best performing one (right), the BDTD, are shown.}
\label{fig:ROC_BDTD}
\end{figure*}

For every event and particular choice of jets, we combine all Signal (Background) distributions to compute an overall Signal (Background) likelihood. By comparing the two likelihoods, a single  classifier value is calculated, one for each multivariate algorithm tested, i.e., a simple BDT, a BDT with a prior step of input feature De-correlation (BDTD), the Fisher method, a likelihood test, etc. As observed in previous analyses, the particular shape of the signal distributions and the way right and wrong combinations are defined make the BDT family the most effective class of classifiers. Among them, the BDTD performs best, giving the largest Area Under the Curve (AUC) and achieving an almost complete separation between right and wrong combinations.
The value of each classifier measures how likely the particular choice of jets matches the kinematics of our signals, at parton level. In Figure~\ref{fig:ROC_BDTD} we show the Receiver Operating Characteristic (ROC) curve for several multivariate classifiers (left) and the best performing one (right), the BDTD. From the ROC curve we see that, if we choose a working point where the BDTD classifier is required to identify the correct jet-parton matching with a probability (signal efficiency) of about 95\%, it rejects incorrectly those (background rejection) with essentially the same probability. This is clearly visible in Figure~\ref{fig:ROC_BDTD} (right). If only positive values of the BDTD classifier are chosen, correct matches are predominately selected (blue area above zero) with a minimal contamination from incorrect ones (red area above zero). Once the best choice of jets is achieved, we proceed to fully reconstruct the events using a kinematic fit, described in what follows.

\vspace*{2mm}
{\it Kinematical Reconstruction}
\vspace*{2mm}

The reconstruction of the $t\bar{t}H_{2,3}$ system is in practice determined by the ability of reconstructing the four-momenta of the two (undetected) neutrinos that originated in the top quark decays. This is accomplished by imposing energy-momentum conservation and mass ($W^\pm$ and $t$) constraints to the events. In order to avoid fixed values for the $W^\pm$ bosons and top quarks, we randomly sample mass values according to the $W^\pm$ bosons and top quark Breit-Wigner distributions, at parton level. Firstly, a two-dimensional probability density function ({\it p.d.f.}) for $m_t$ versus $m_{\bar{t}}$ is used to generate random mass values for the top quarks. Secondly, $m_{W^+}$ and $m_{W^-}$ are generated from the two-dimensional {\it p.d.f.}'s of ($m_t,m_{W^+}$) and ($m_{\bar{t}},m_{W^-}$), respectively, such that  physical correlations are preserved in the reconstruction. 
We apply the following mass constraints to the $t\bar{t}$ system:
\begin{eqnarray}
\label{equ:a0}
\ensuremath{( p_{\ell +(-)} + p_{\nu(\bar{\nu})} )^{2} &=& m_{W^{+(-)}}^{2}}, \\
\label{equ:a1}
\ensuremath{( p_{W^{+(-)}} + p_{b(\bar{b})} )^{2} &=& m_{t(\bar{t})}^{2}},
\end{eqnarray}
\noindent
where $p_{b(\bar{b})}$, $p_{\ell+(\ell-)}$ and $p_{\nu(\bar{\nu})}$ correspond to the four-momenta of the $b$-jets, the charged leptons and neutrinos from the $t$($\bar{t}$) decays, respectively. The four-momenta of the $W^{+(-)}$ bosons are represented by $p_{W^{+(-)}}$. We assume that the undetected neutrinos are fully responsible for the Missing Transverse Energy (MET), i.e.,
\begin{eqnarray}
\label{equ:a2}
\ensuremath{p_{x(y)}^{\nu} + p_{x(y)}^{\bar{\nu}} &=& \slash\kern-.6emE_{x(y)}},
\end{eqnarray}
where $\ensuremath{\slash\kern-.6emE_{x(y)}}$
represents the $x$ ($y$) components of the MET. If a solution is not found for the particular choice of top quark and $W^\pm$ boson masses, the generation of mass values is repeated, up to a maximum of 500 tries, until at least one solution is found. If still no solution is found, the event is discarded as not being compatible with the topology under study. 
For a given event, Eqs.~\eqref{equ:a0}--\eqref{equ:a2} may give more than one possible solution. For each one, we calculate a likelihood of being consistent with a $t\bar{t}H_{2,3}$ dileptonic event. Its value is computed as the product of {\it p.d.f.} built from the $p_T$ distributions of the neutrino, top quarks and $t\bar{t}$ systems as well as the two dimensional {\it p.d.f.} of the top quark masses. The solution with the highest value is chosen as the correct one. It is worth mentioning that, whenever we take the jets correctly matched to their corresponding partons in $t\bar{t}H_{2,3}$ events, the kinematical fit finds at least one solution in 60-70\% of the cases. 
Moreover, since both $H_2$ and $H_3$ are yet to be discovered, no constraints are imposed on the mass of the $b\bar{b}$ pairs originating from their decays, even though the jet–parton matching correctly assigns the jets to the $H_{2,3}$ decays. This is a key difference from a traditional SM Higgs analysis, where the mass of the $b\bar{b}$ system can be fixed to the known value of the SM Higgs boson mass (about 125 GeV). This distinction reinforces our strategy of focusing on the properties of the $t\bar{t}$ system until the $H_{2,3}$ states are discovered. Once their masses are established, such constraints may then be applied.

\begin{table}[h]
\renewcommand{\arraystretch}{1.3}
\centering
\hspace*{-0.25truecm}
\begin{tabular}{l|c|c|c|c}
    \toprule
    &   \textbf{Pre-Sel.} & \multicolumn{3}{c}{\textbf{Final Selection}} \\[-1mm]
    \cmidrule(lr){2-5}
    &   ~$N_{\rm jets} \ge$~4~   &   Kinematic  &   $m_Z$   &   $N_b$  	\\[-1mm]
    &   ~$N_{\rm leptons} \ge$~2~    &    Fit       &    cut    &   = 4     \\[-1mm]
    \midrule  
     $t\bar{t}$+$c\bar{c},t\bar{t}$\,+\,light flavors         &    1122.78              &    672.94     &    579.01  &   2.19     \\         
     $t\bar{t}+Z$ ($Z\rightarrow\nu\bar{\nu}$) &    	0.40    	       &    0.16      &    0.13   &   0.00     \\     
     $t\bar{t}$+$V~(V$=$W^\pm,Z)$    			      &    8.23    	       &    4.62      &    3.98   &   0.07     \\ 
     $t\bar{t}+H_1$                            &    	4.13    	       &    2.38      &    2.06   &   0.27     \\     
     Single $t$    						      &    33.38      	       &    15.31     &   13.31   &   0.00     \\
     $W^\pm$+jets (with $W^\pm b\bar{b}$)    			  &   $\le 0.001$     	   &     0.00     &    0.00   &   0.00     \\              
     $Z$+jets (with $Z b\bar{b}$)                 &    94.47     	       &    43.60     &    2.14   &   0.00     \\              
     Di-boson ($W^+W^-$,			      &                  &        &      &       \\      
        $W^\pm Z$ and $ZZ$)   			      &    36.75               &    16.12     &    1.48	  &   0.00     \\         
    \midrule  
     Total background    		              &   1300.14    	       &    755.13     &    602.10  &   2.53      \\ 
    \midrule  
     $t\bar{t}H_2$    					      &      0.087       	   &    0.049     &    0.042  &      0.006  \\       
     $t\bar{t}H_3$    					      &      0.063      	   &    0.035     &    0.031  &      0.004  \\ 
    \bottomrule
\end{tabular}
\caption{Expected cross sections (in fb) after pre-selection (Pre-Sel.) and Final Selection (for several applied criteria) for dileptonic signals and SM background events at the LHC.}
\label{tab:cutflow}
\end{table}

\begin{table}[htbp]
    \centering
    \begin{tabular}{l|c|c}
        \toprule
        \textbf{MC Parameter} & \textbf{\boldmath$t\bar{t}H_2$} & \textbf{\boldmath$t\bar{t}H_3$} \\
        \midrule
        Yukawa couplings (norm. to SM) & $-0.254$ & $-0.197$ \\
        Reference cross section [pb] & $4.85 \times 10^{-4}$ & $3.39 \times 10^{-4}$ \\
        Final selection efficiency ($\epsilon$) & $0.013$ & $0.013$ \\
        \bottomrule
    \end{tabular}
    \caption{MC parameters used to generate the signal events and final selection efficiency.}
    \label{tab:mc_params}
\end{table}

\vspace*{2mm}
{\it Final Selection and Cross Section Yields}
\vspace*{2mm}

\begin{figure*}
\begin{center}
\begin{tabular}{ccc}
\hspace*{-5mm}\includegraphics[height=7.5cm]{ 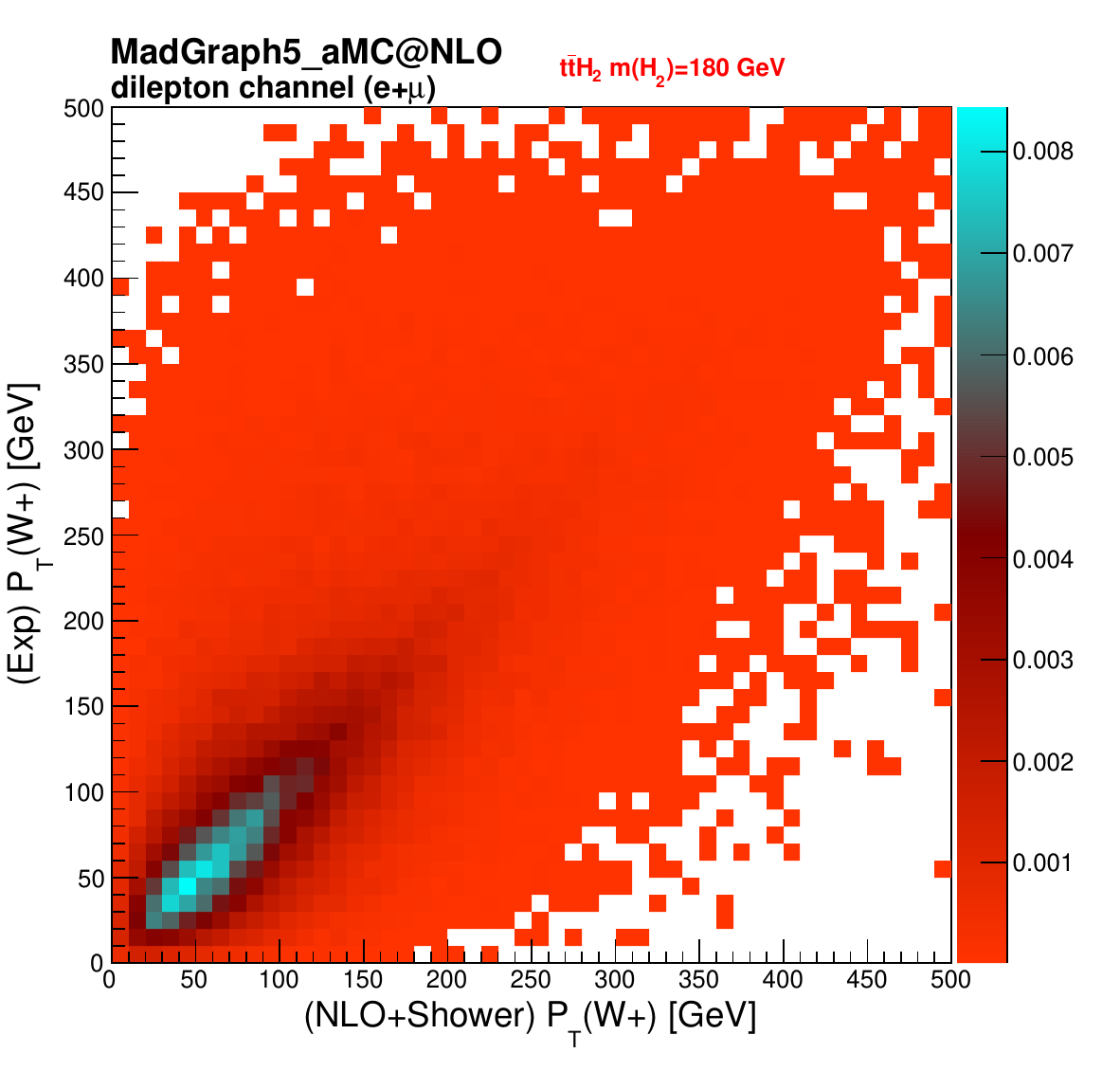}
\hspace*{5mm}\includegraphics[height=7.5cm]{ 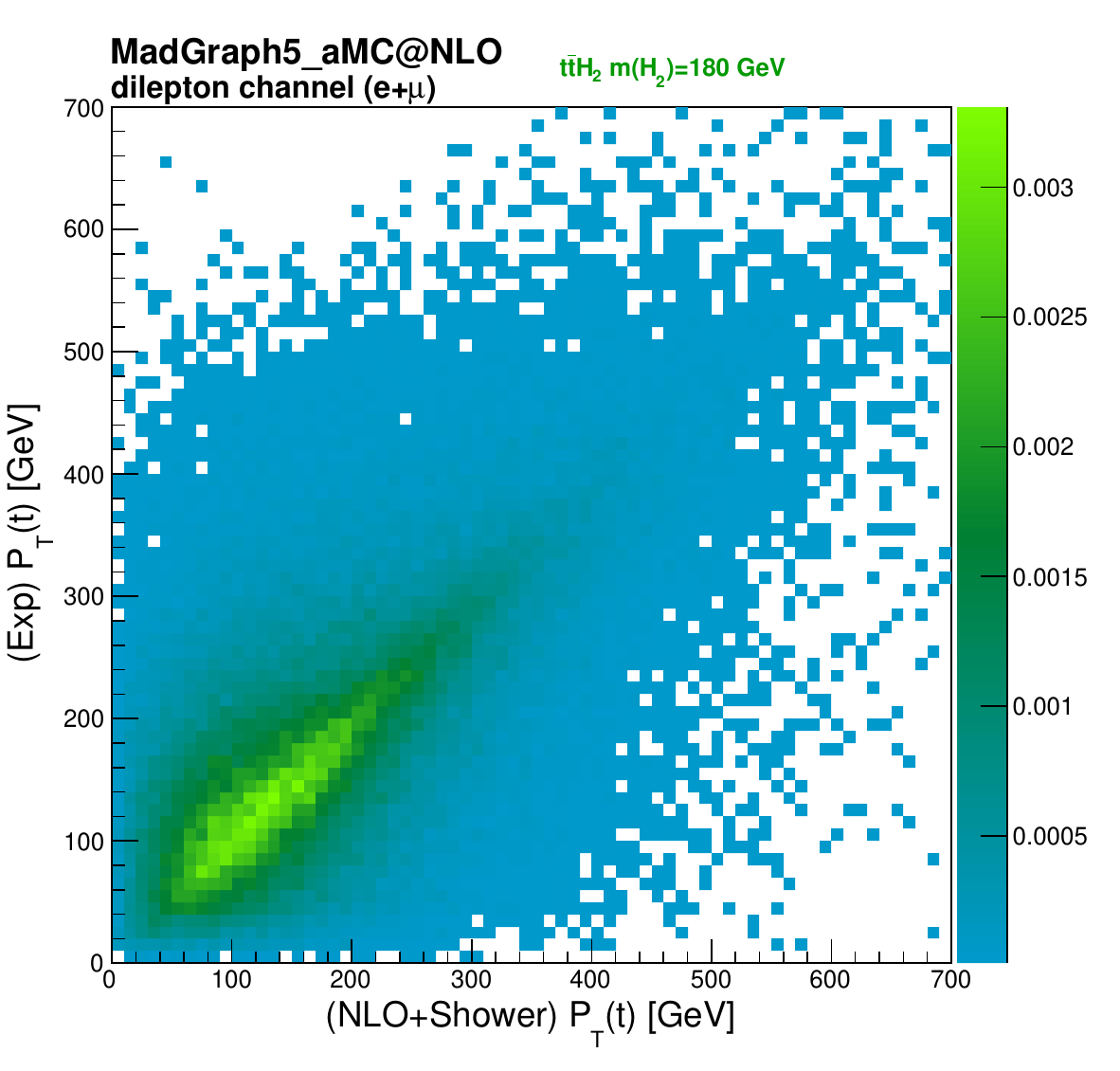}\\
\hspace*{-5mm}\includegraphics[height=7.5cm]{ 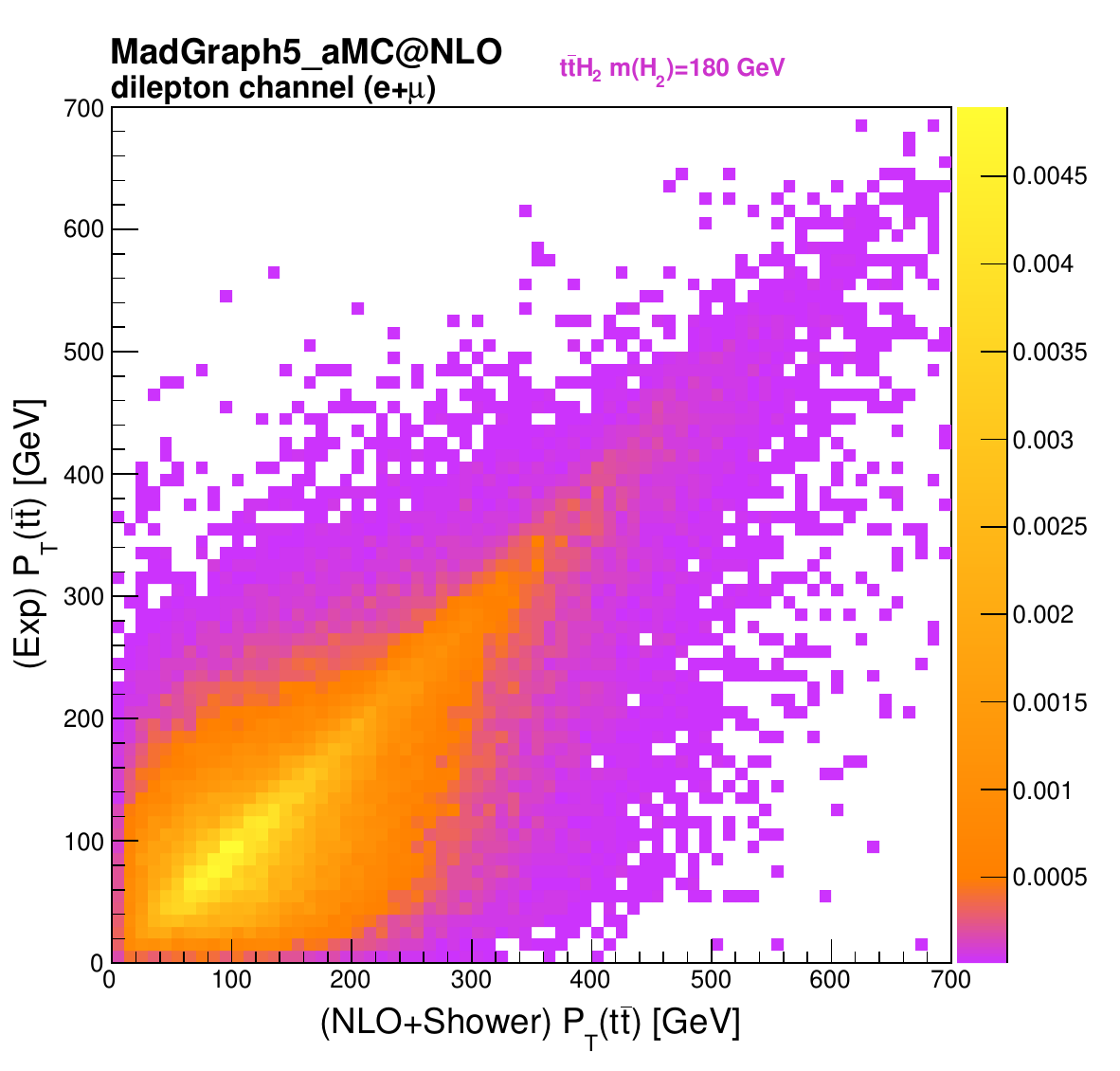}
\hspace*{5mm}\includegraphics[height=7.5cm]{ 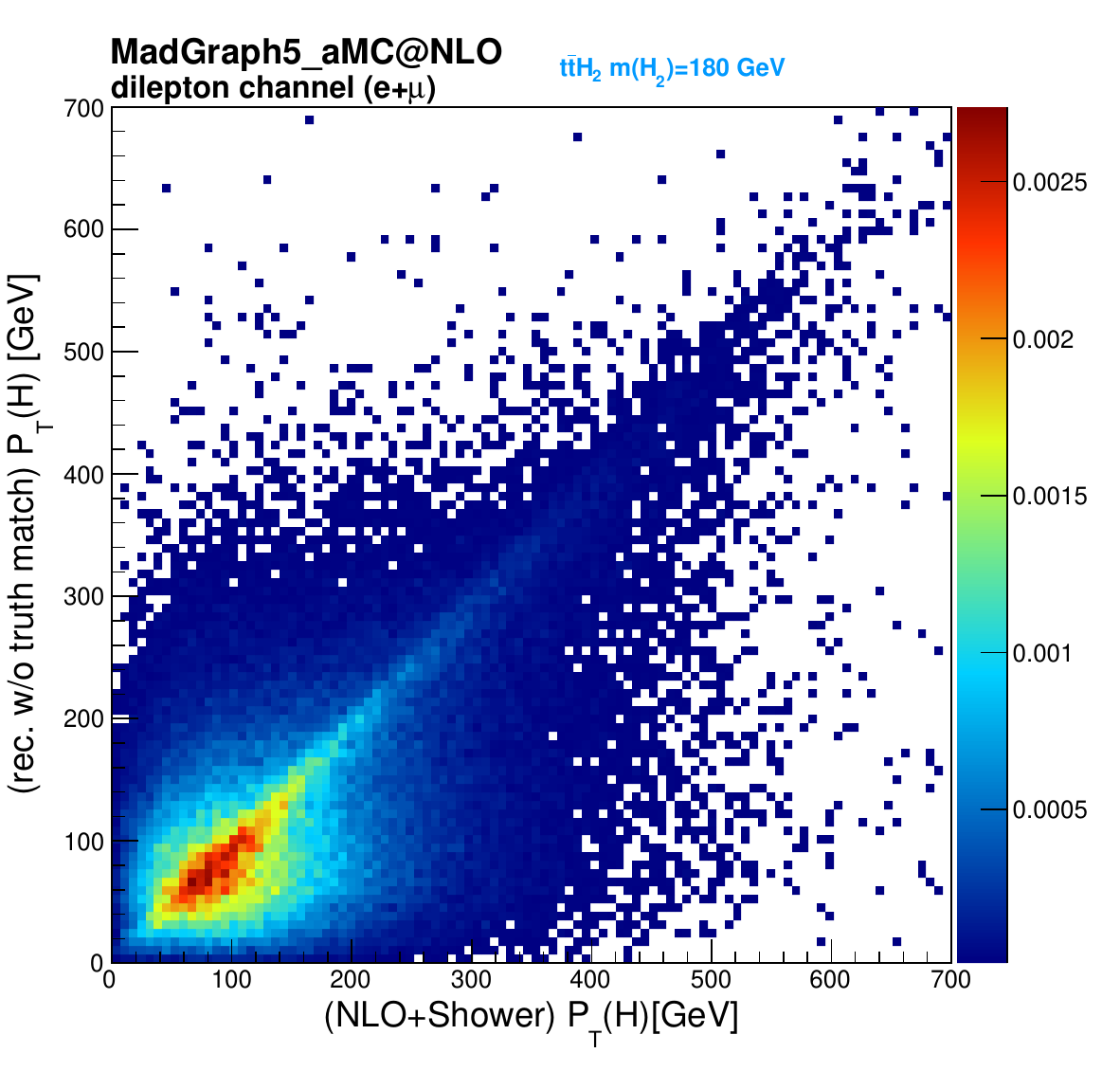}
\end{tabular}
\end{center}
\caption{Two-dimensional $p_T$ distributions for $t\bar{t}H_2$ events. The horizontal axes show variables evaluated at parton level. The vertical axes represent the corresponding variables after kinematic reconstruction at detector level. We show the distribution for the $W^+$ boson (upper-left), the $t$-quark (upper-right), the $t\bar{t}$ system (bottom-left) and $H_2$ boson (bottom-right). A clear correlation between parton and reconstructed levels is found. The same behaviour is observed for $t\bar{t}H_3$ events.}
\label{fig:Kinematic_Rec_ttH}
\end{figure*}

\begin{figure*}
\begin{center}
\begin{tabular}{ccc}

\hspace*{-5mm}\includegraphics[height=7.5cm]{ 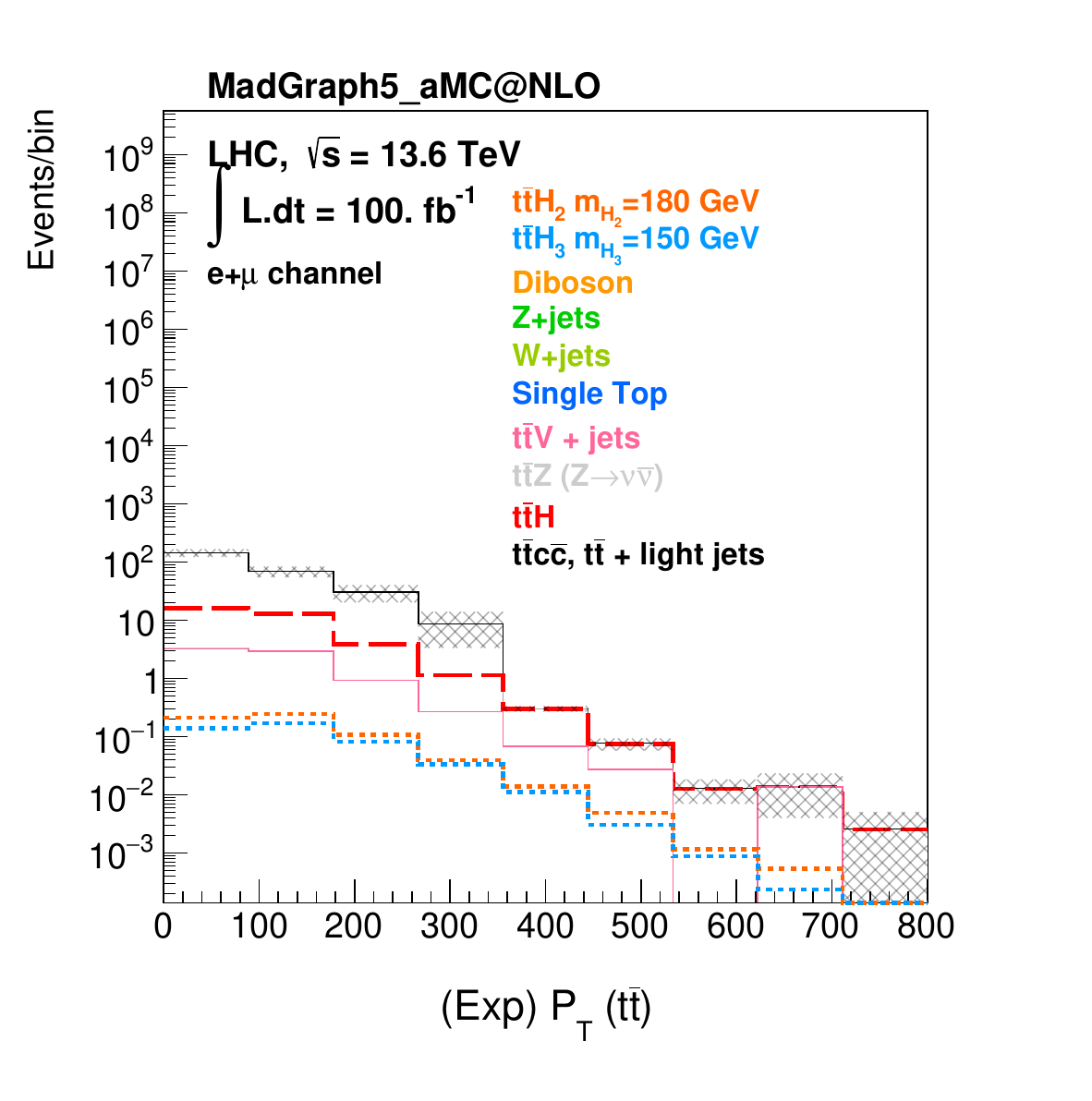}
\hspace*{5mm}\includegraphics[height=7.5cm]{ 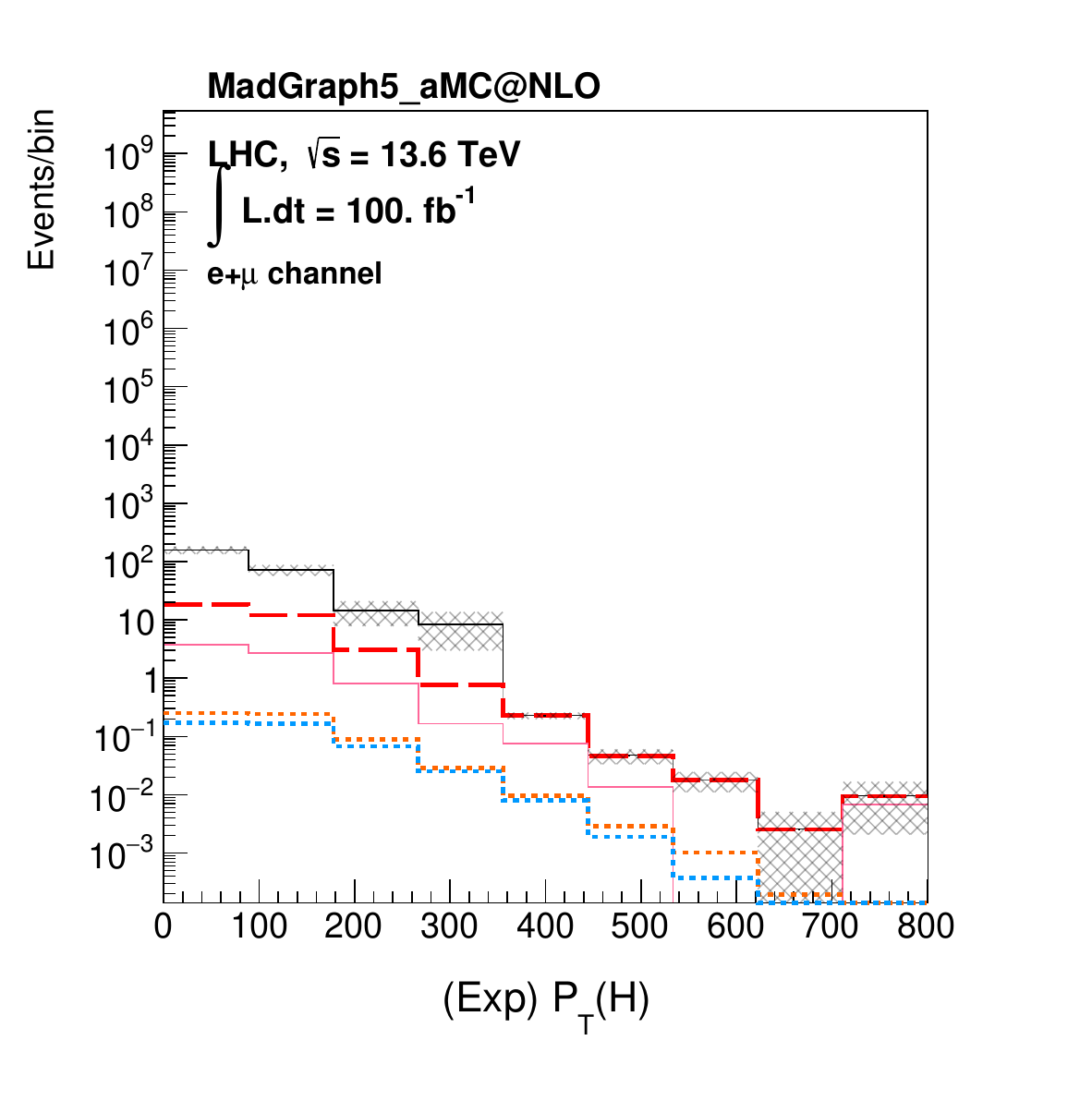}\\
\hspace*{-5mm}\includegraphics[height=7.5cm]{ 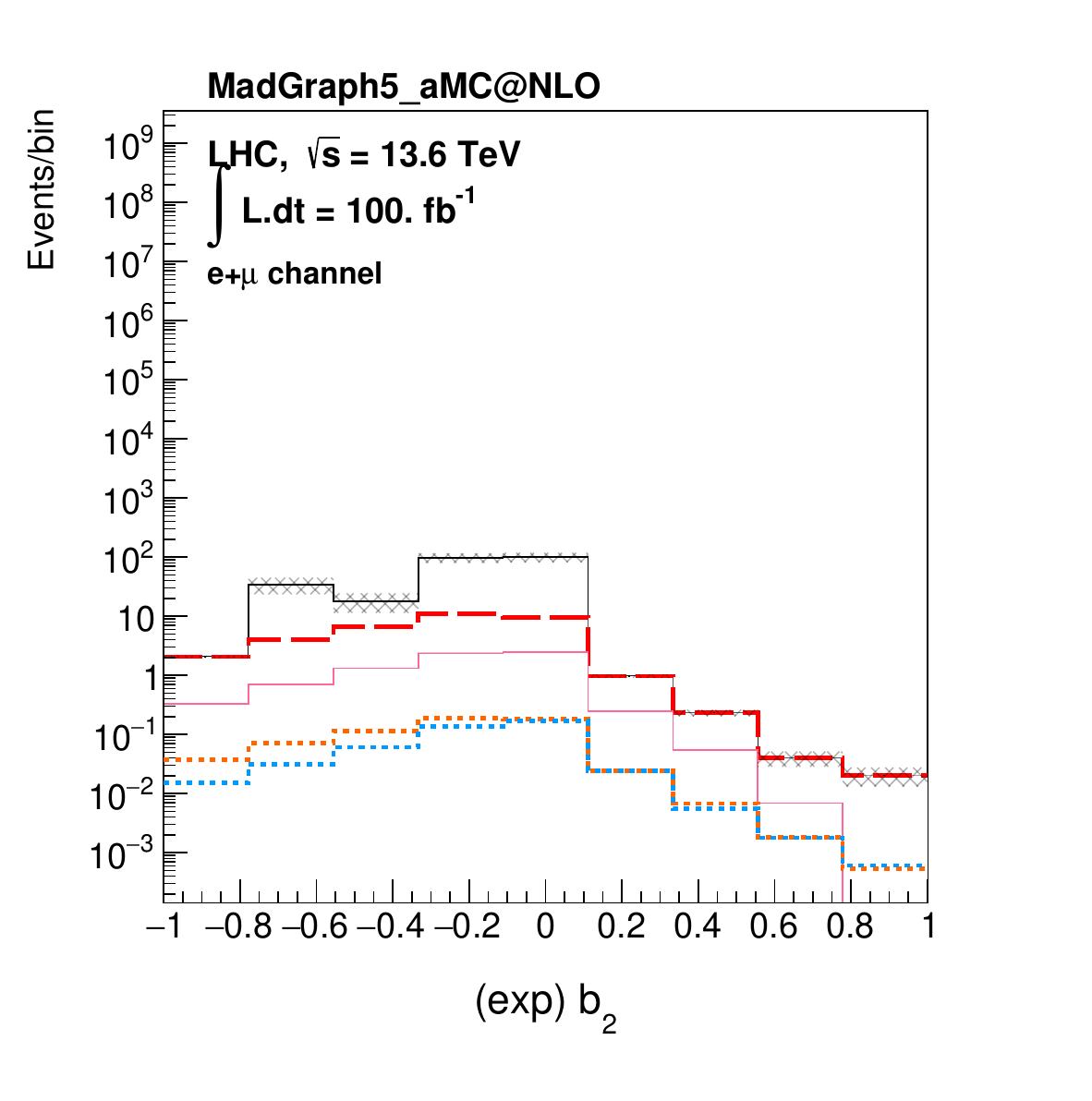}
\hspace*{5mm}\includegraphics[height=7.5cm]{ 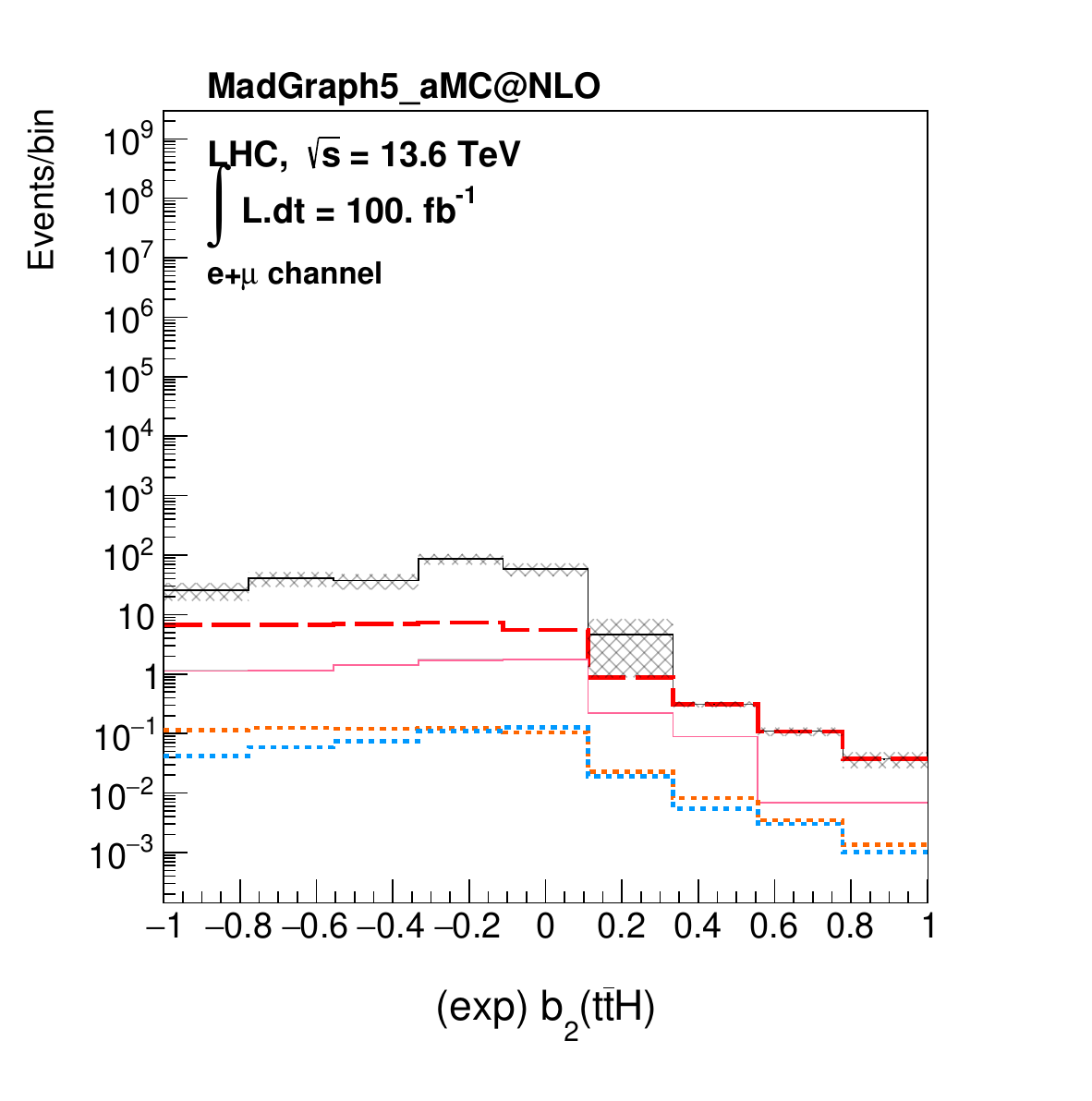}\\
\hspace*{-5mm}\includegraphics[height=7.5cm]{ 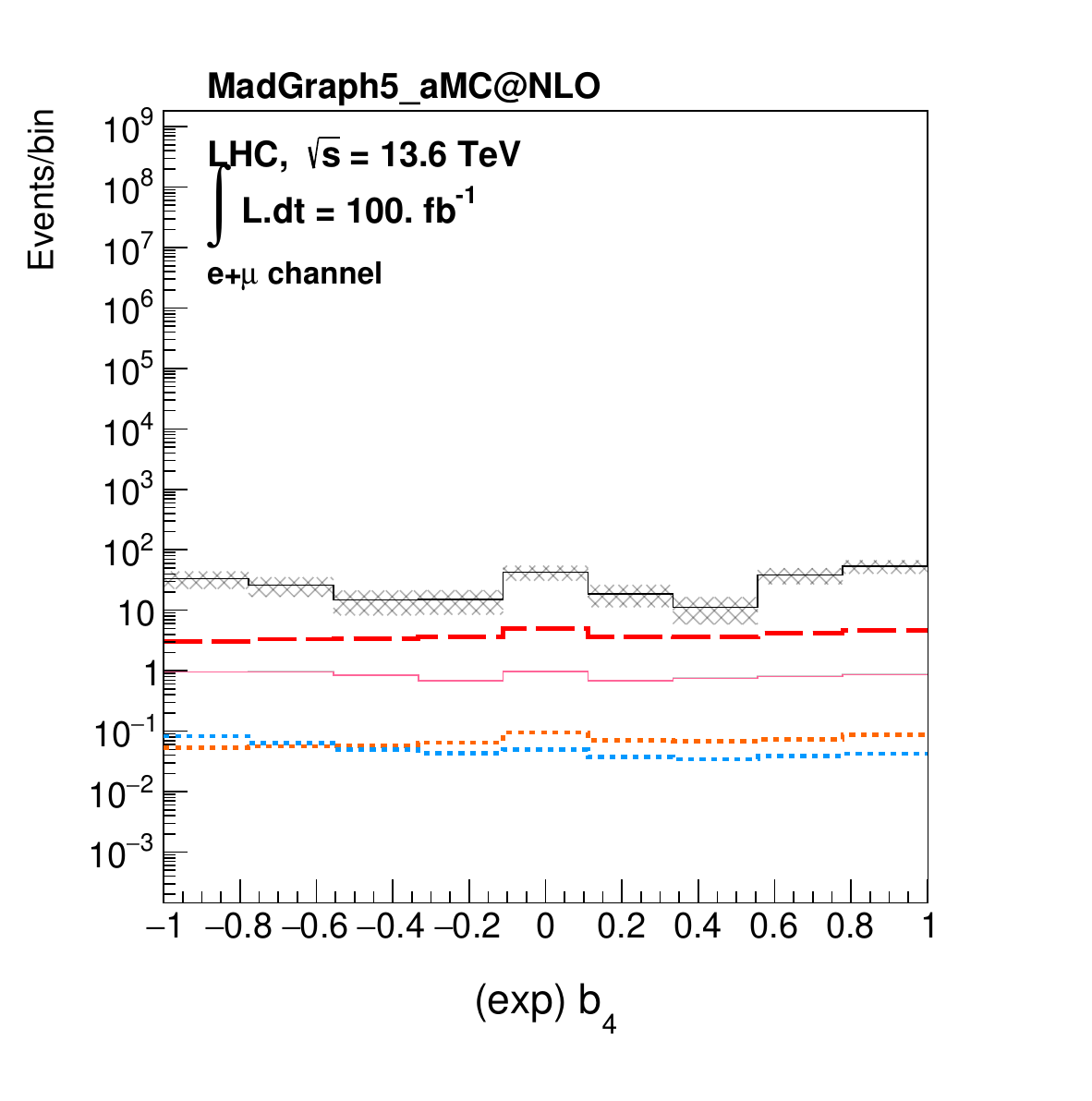}
\hspace*{5mm}\includegraphics[height=7.5cm]{ 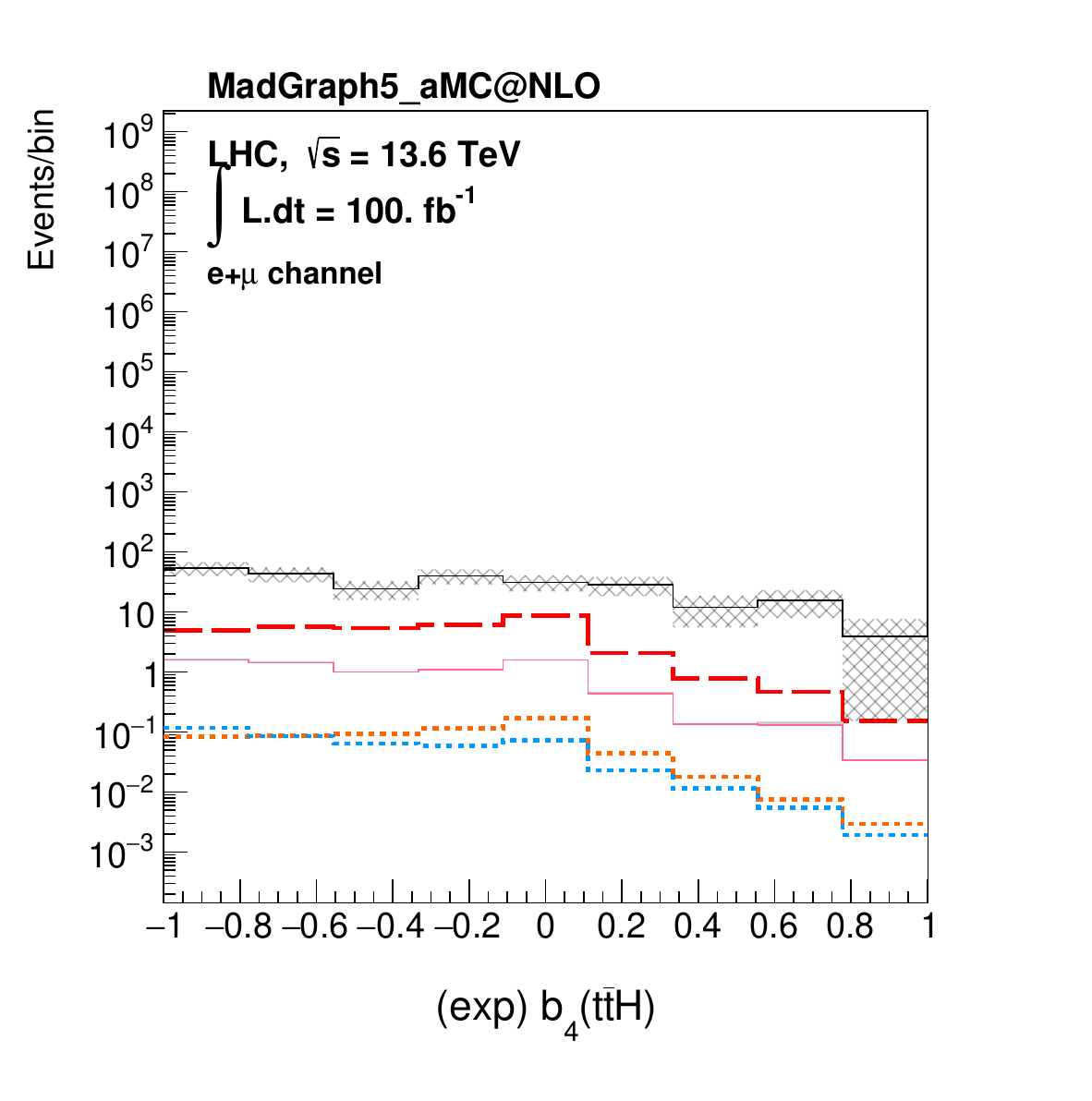}
\end{tabular}
\end{center}
\caption{(Upper-left) The transverse momentum of the $t\bar{t}$ system, $p_T(t\bar{t})$, and (upper-right) Higgs boson, $p_T(H)$, both evaluated in the LAB frame. The $b_2$ (middle) and $b_4$ (bottom) variables are also shown in the LAB (left) and CoM (right) frames. The distributions are represented for a reference luminosity of $L$=100~fb$^{-1}$.}
\label{fig:Angular_HeavyHiggs}
\end{figure*}

To increase the sensitivity of the analysis, we apply additional selection criteria to pre-selected events, profiting from the kinematic reconstruction. In the Final Selection, we only accept events if at least one solution of the kinematic fit was found (see Table~\ref{tab:cutflow}). Moreover, to mitigate the contribution from the SM backgrounds involving $Z$ bosons that decay to oppositely charged leptons ($Z\rightarrow\ell^+\ell^-$), we only accept events if the invariant mass of the leptons ($m_{\ell^+\ell^-}$) is outside a window of 10~GeV size around the $Z$ boson mass ($m_Z$), i.e., $|m_{\ell^+\ell^-}-m_Z|>10$~GeV (see Table~\ref{tab:cutflow}, {``$m_Z$ cut''}). We mitigate the contribution from irreducible top quark backgrounds by accepting events with exactly 4 $b$-tagged jets (see Table~\ref{tab:cutflow}, ``$N_b$=4''), two from the top quarks and the other two from the Higgs bosons decays. In Table~\ref{tab:cutflow} we show the expected cross sections in fb, at several stages of the event selection, for dileptonic signal and SM backgrounds. The final selection efficiencies for the signals, after all cuts are applied,  are shown in Table~\ref{tab:mc_params}, for the particular choice of parameters used in our reference analysis.
In Figure~\ref{fig:Kinematic_Rec_ttH} we show examples of two-dimensional $p_T$ distributions of the $W^+$ (upper-left), the top quark (upper-right), the $t\bar{t}$ system (bottom-left) and the $H_2$ boson (bottom-right), after kinematic reconstruction of $t\bar{t}H_2$ events, $Z$-boson window selection and requiring 4 $b$-jets in the events. The correlation between the parton level ($x$-axis) and reconstructed ($y$-axis) $p_T$ distributions, is clearly visible. The same behavior is observed for the $t\bar{t}H_3$ signals. In Figure~\ref{fig:Angular_HeavyHiggs}, we show the reconstructed transverse momentum of the $t\bar{t}$ system (upper-left) and Higgs boson (upper-right), both evaluated in the LAB frame, for a reference luminosity of $L$=100~fb$^{-1}$, and after our Final Selection. The background composition is also shown. While the $t\bar{t}$+jets process largely dominates the SM background, in particular, in regions of low $p_T$, signals tend to extend  also to regions of higher $p_T$ values.  Quite importantly, here, it is worth noting that the distributions are rather similar for both the $t\bar t$ and Higgs systems, thereby signaling again that it may be worthwhile entertaining an inclusive analysis on the latter rather than engaging in its reconstruction. 

In Figure~\ref{fig:Angular_HeavyHiggs}, we also show the  distributions for $b_2$ (middle plots) and $b_4$ (lower plots) in the LAB (left) and CoM (right) frames, for the Higgs boson signals and the dominant background processes. As previously, the distributions are represented for a reference luminosity of $L$=100~fb$^{-1}$ and after our Final Selection. 
Here, the normalization of the $H_{2,3}$ signals is important for representation purposes, as it allows them to stand out above background fluctuations. This makes the shape of the distributions easier to perceive, enabling us to see not only how the signals differ from one another but also how they differ from the SM backgrounds.
When moving from the LAB to the $t\bar{t}H_{2,3}$ reference systems, clear differences are observed in the shape of the signals, however, these can precisely be modeled by MC. Again, we can see the discriminant power of these variables, between the CP-even and CP-odd Higgs states, which clearly persists after a full reconstruction of the final state, thereby signaling the solidity of an approach exploiting these angular distributions.

\subsection{Sensitivity to the Higgs Masses} 
\begin{figure*}
\begin{center}
\begin{tabular}{ccc}
\hspace*{-5mm}\includegraphics[height=6cm]{ 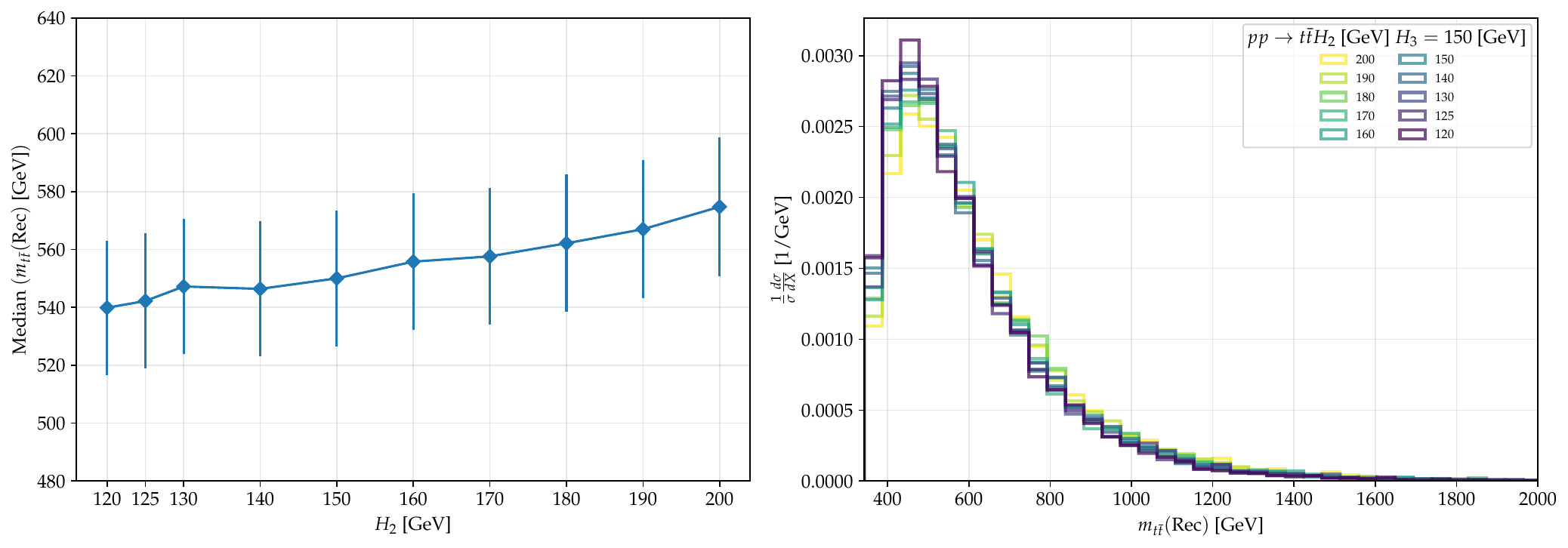}\\
\hspace*{-5mm}\includegraphics[height=6cm]{ 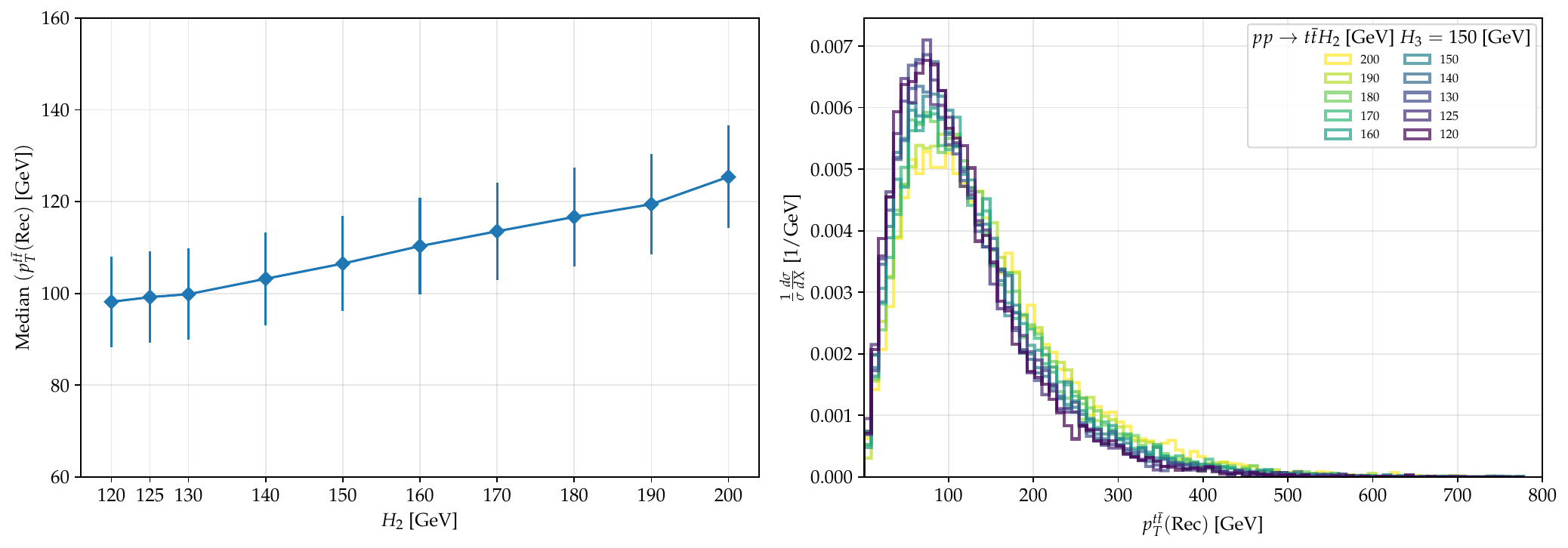}\\
\end{tabular}
\end{center}
\caption{Sensitivity of key kinematic observables to the mass of the extra CP-even Higgs boson ($m_{H_2}$) in $pp \to t\bar{t}H_2$ events, after event selection and full kinematic reconstruction of the $t\bar{t}H$ system. (Left). As an example, the 50$^{th}$ percentile of the $m_{t\bar{t}}$ (top) and $p_T^{t\bar{t}}$ (bottom) distributions are shown as  functions of $m_{H_2}$. (Right). The corresponding distributions normalized to the cross section for several  mass values, illustrating the sensitivity of the observables to the mass gap $|m_{H_2}-m_{H_3}|$.}
\label{fig:exp_mttbar_ptttbar_quartiles}
\end{figure*}

\begin{figure*}
\begin{center}
\begin{tabular}{ccc}
\hspace*{-5mm}\includegraphics[height=5.5cm]{ 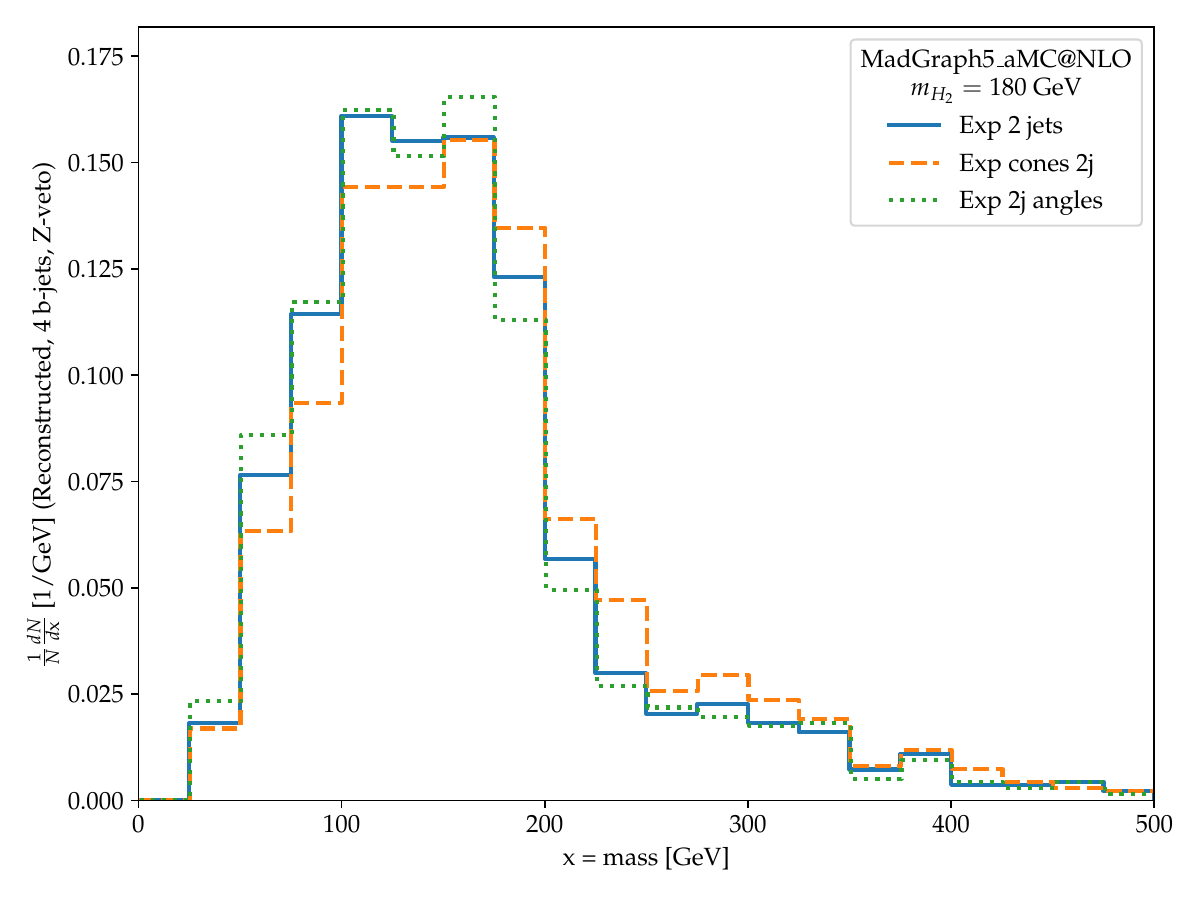}
\hspace*{-0mm}\includegraphics[height=5.5cm]{ 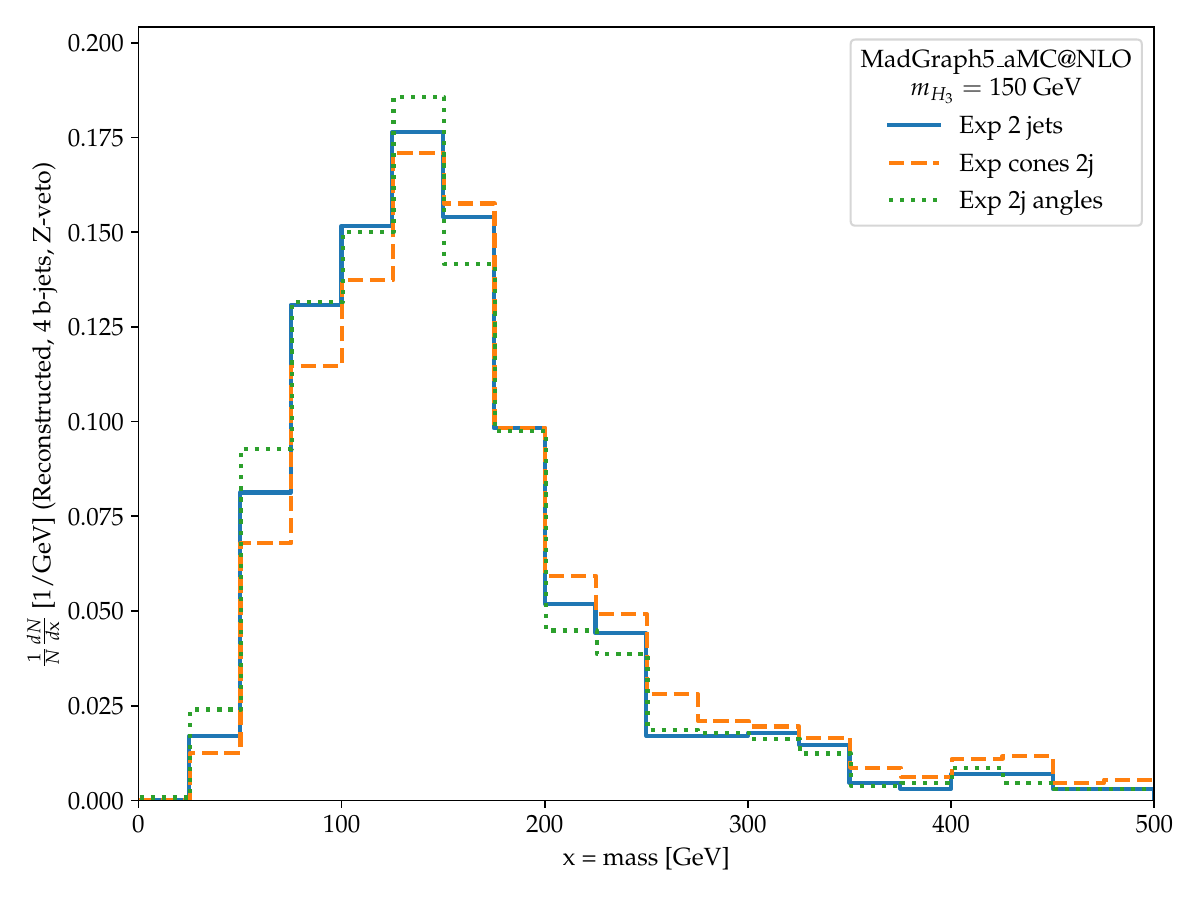}\\
\end{tabular}
\end{center}
\caption{Left(Right): Direct reconstruction of the mass of the additional Higgs state $H_{2(3)}$ from the $b\bar{b}$ system, {{using $m_{H_{2(3)}}=180(150)$ GeV}}.}
\label{fig:bbar_rec}
\end{figure*}

One of the questions we address in this paper is whether the $m_{t\bar{t}}$ and $p_T^{t\bar{t}}$ variables retain any of the parton level sensitivity to the mass of the recoil system even after event selection, i.e., upon matching at detector level the right jets to their mother particles (the top quarks and Higgs bosons) and following full kinematic reconstruction of the $t\bar{t}H_{2,3}$ system. In Figure~\ref{fig:exp_mttbar_ptttbar_quartiles} (right) we show the $m_{t\bar{t}}$ (top-right) and $p_T^{t\bar{t}}$ (bottom-right) variables after all this has been enforced. 
As an example, the 50$^{th}$ percentile of the $m_{t\bar{t}}$ (top-left) and $p_T^{t\bar{t}}$ (bottom-left) distributions are shown as a function of $m_{H_2}$. As we saw already at parton level in Figure~\ref{fig:mttbar_ptttbar_quartiles} (left), indeed, even after applying the full analysis chain, the variables still show a clear dependence upon the mass of the recoiling system. The direct reconstruction of the $H_{2,3}$ masses using the invariant mass of the $b\bar{b}$ system from the $H_{2,3}$ decays, including not only the energy-momentum information of the jets from the $b$-quarks but also considering the whole energy-momentum within a cone of $\Delta R\le0.4$ around both jets, is shown in Figure~\ref{fig:bbar_rec} (with label {{``Exp cones 2j''}}). Here we see that, although the distributions show peaks at the expected mass values, their resolution is not great, even when using alternative methods that includes the information of only the two jets (labeled {{``Exp 2 jets''}}) or the most energetic jet and their angles (assuming they both came form the same Higgs boson decay, labeled {{``Exp 2j angles''}}), making a precise mass determination very challenging. The fact that we may have the possibility of accessing the mass of the recoiling system, in addition to the direct mass measurements (with the described limitations), is indeed quite interesting. Although the attempt to make such a mass extraction seems doable, it stays largely outside the scope of this paper, and may be revisited in future works.


\begin{figure*}
\begin{center}
\begin{tabular}{ccc}
\hspace*{-5mm}\includegraphics[height=6.5cm]{ 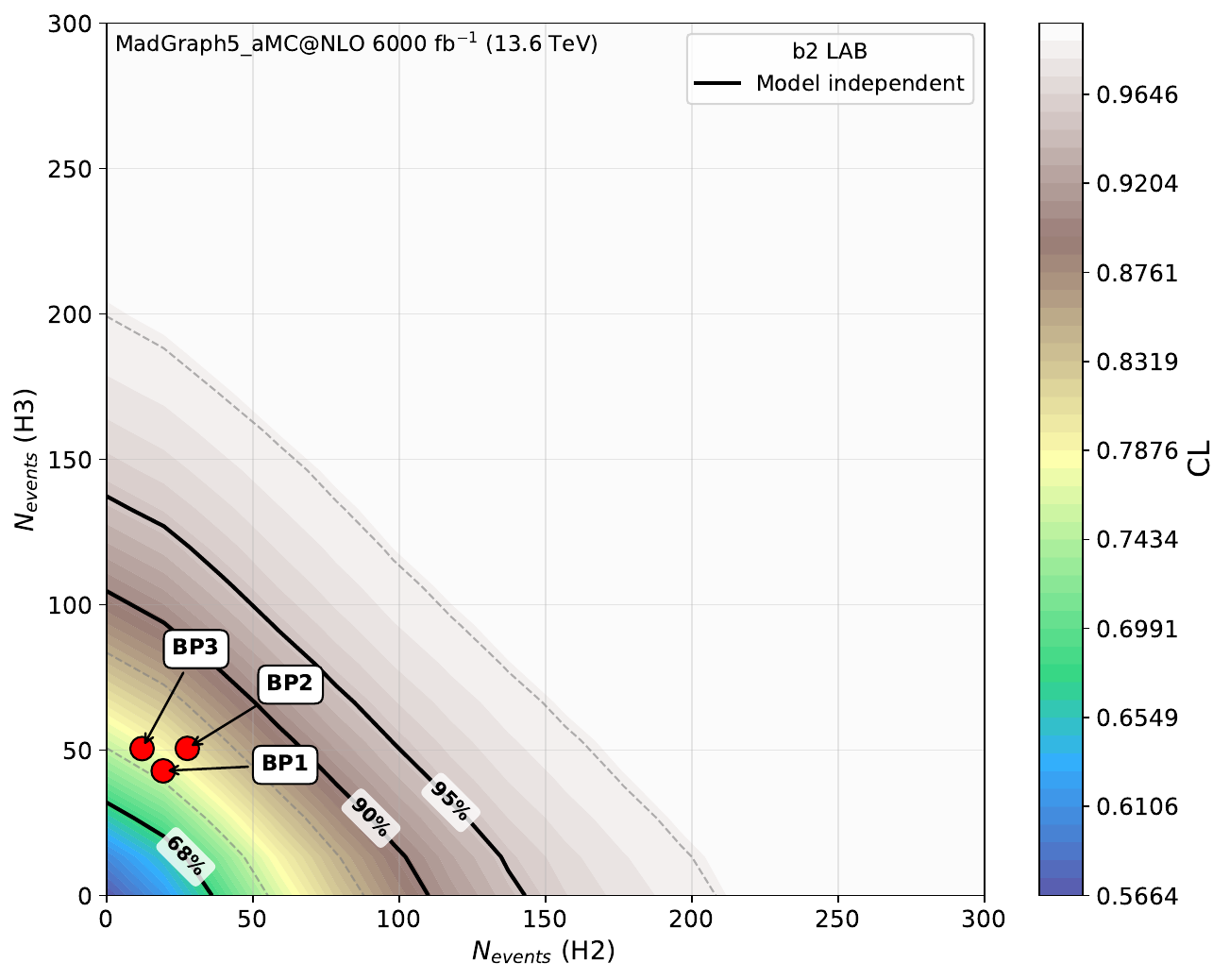}
\hspace*{-1.25mm}\includegraphics[height=6.5cm]{ 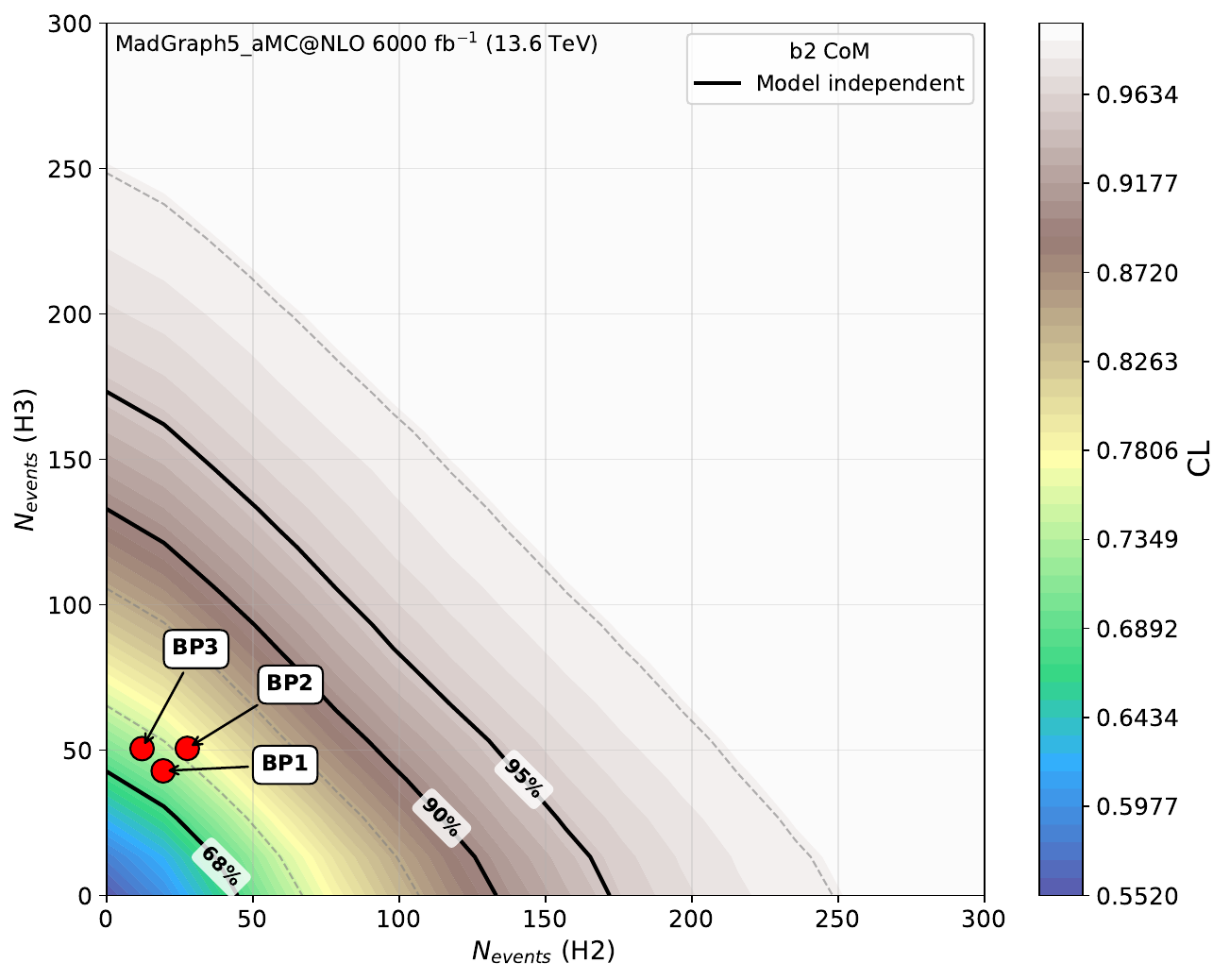}\\
\hspace*{-5mm}\includegraphics[height=6.5cm]{ 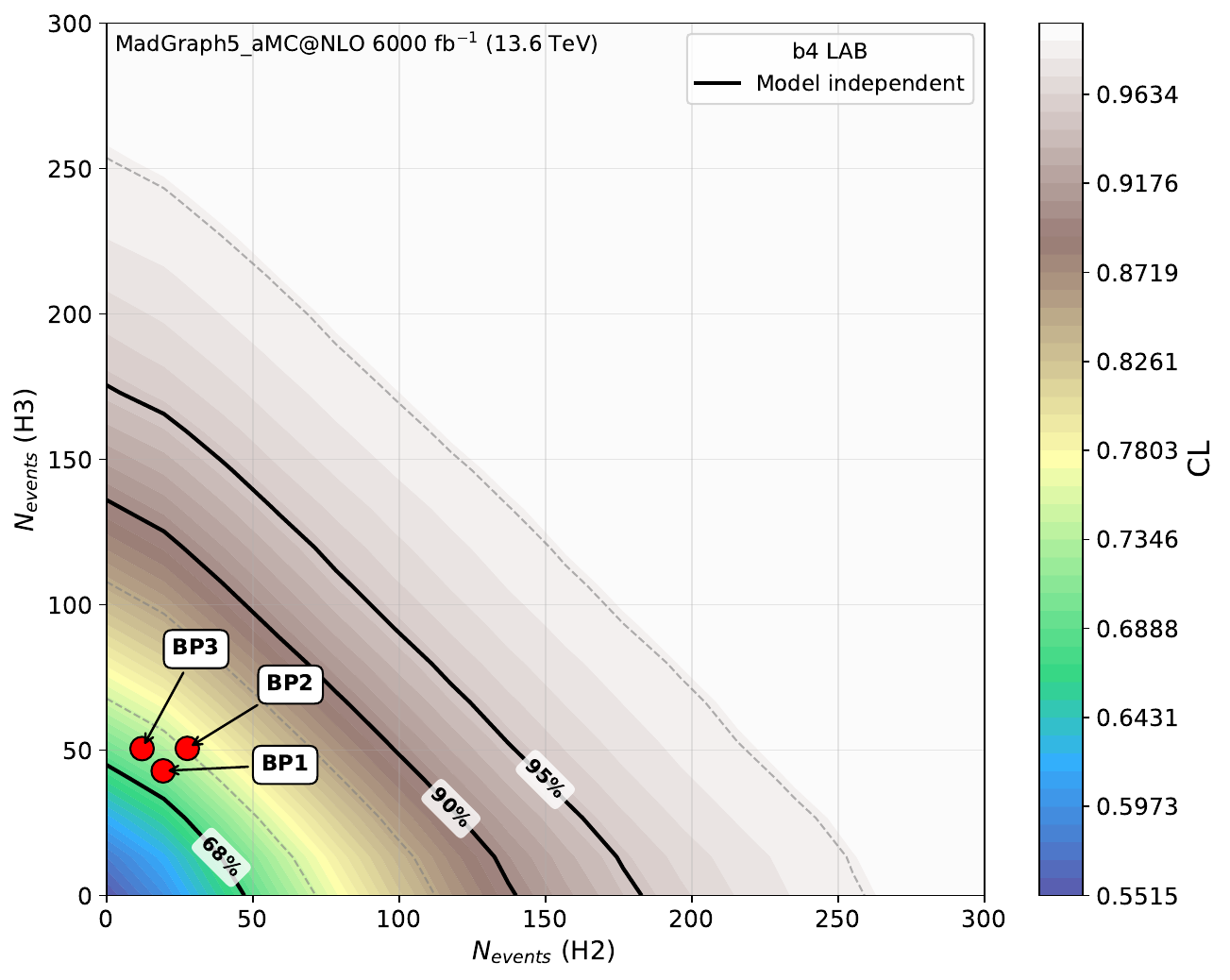}
\hspace*{-1.25mm}\includegraphics[height=6.5cm]{ 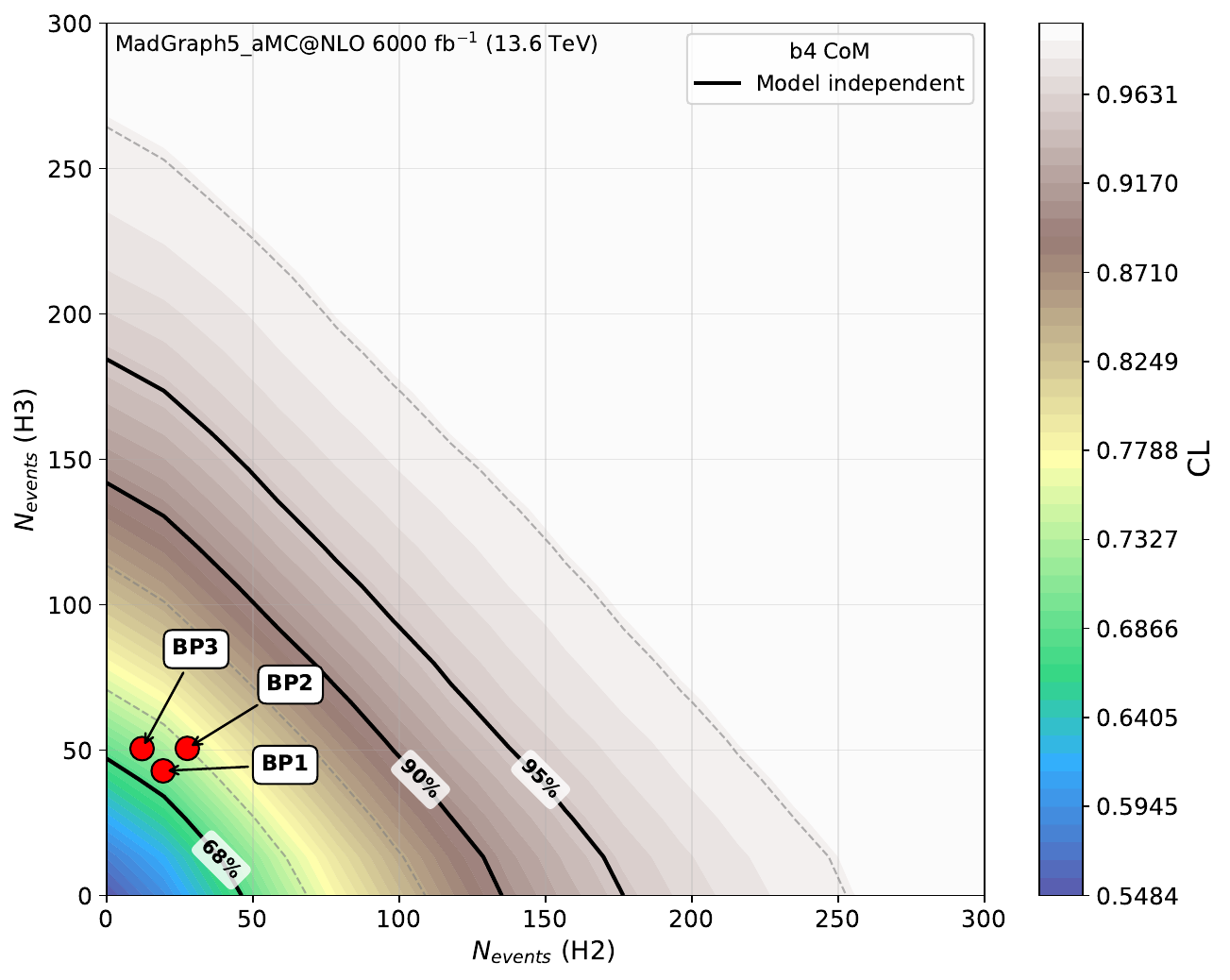}
\end{tabular}
\end{center}
\caption{Model independent exclusion limits on the event rates of the extra Higgs bosons ($H_{2,3}$) 
computed against the SM (with a 125~GeV Higgs boson) as null hypothesis for the HL-LHC ($L$=6000~fb$^{-1}$) as obtained from  the $b_2$ (top) and $b_4$ (bottom) observables, evaluated in the LAB (left) and CoM (right) frame. 
{{The sampled data points from the C2HDM (with $m_{H_{2,3}}< 2m_{t}$) are also shown as red points.}} }
\label{fig:CLs_HeavyHiggs}
\end{figure*}

\subsection{Sensitivity to the Higgs CP State}
The same question can be asked in the case of the variables displaying sensitivity to the CP quantum number of the system recoiling against the $t\bar t$ pair. 
We thus focus again on the $b_2$ and $b_4$ angular distributions in both the LAB and  CoM reference frames to estimate expected exclusion limits for two signal scenarios. These  involve the new  Higgs bosons, $H_2$ and $H_3$, with different masses and CP properties, i.e., CP-even and CP-odd, respectively. We assume that the SM background, including the contribution from the SM Higgs boson,  is well understood and fully modeled using our MC simulations. We consider the following case: we test how well, i.e., with which Confidence Level (CL), the SM, considered as the null hypothesis (H0), can be excluded in favor of a signal hypothesis (H1) with a mixture of $t\bar{t}H_2$ and $t\bar{t}H_3$ signals.

In Figure~\ref{fig:CLs_HeavyHiggs}, the additional Higgs ($H_{2,3}$) exclusion limits are shown for such scenario. 
Limits are represented for the full luminosity ($L$=6000~fb$^{-1}$, integrated across ATLAS and CMS) of the High-Luminosity phase of the LHC (HL-LHC) \cite{Gianotti:2002xx,Apollinari:2015wtw}, in the LAB (left) and CoM (right) frames. Minor differences (at the level of few~\%) exist between the two sets of  results obtained with both the $b_2$ (upper) and $b_4$ (middle) distributions. The red points correspond to the BPs of the C2HDM listed in Table~\ref{table1} as BP1, BP2 and BP3.
Although no major changes are to be expected when moving from the LAB to the CoM  frame, results evaluated in the latter(former) are normally better for $b_4(b_2)$. 
Altogether these results highlight the sensitivity of $b_2$ and $b_4$ to the CP state of $H_{2(3)}$ (but, recall,  not their masses).  

\subsection{Interpretation}

All results from the previous section are applicable to the generic 2HDM,  for which Yukawa couplings and Higgs masses are all independent input parameters.  For the C2HDM, they are not, as all Langrangian couplings and masses are derived quantities from the fundamental inputs  in the new strong sector, hence, not all the parameter space sampled is accessible to such a BSM scenario. In the last three figures, we display BPs from this BSM construct that have emerged from the previously described scan (red points), since such a sensitivity plots were obtained from $b_2$ and $b_4$, which phenomenology does not depend on the values of $m_{H_{2,3}}$, which change over such a plane for the C2HDM. The emerging message is clear:
while sensitivity to the 2HDM already exists at Run 3 of the LHC (one can simply scale the CL values therein a factor $1/\sqrt{10}$), for the C2HDM this is much reduced. However, at the HL-LHC, sensitivity to both models will exist, albeit limitedly to a handful of BPs in the latter case.  
However, it is worth mentioning that, while this result certainly has some bearing on the 2HDM setup, it is instead very important in the context of the C2HDM, because this scenario predicts values for the Yukawas and masses of the neutral additional Higgses precisely in the regions identified by the present analysis as those to which the HL-LHC will have sufficient sensitivity, in particular, if several variables are combined (a possible future improvement of our analysis), and
for the relevance that this BSM framework may have in remedying the flaws of the SM.

\section{Conclusions \label{sec:conclusions}}
In this paper, we have shown how the study of the top-antitop pair produced in association with an additional spin-0 object $X$ can provide significant insights into both its mass and CP properties, irrespectively of its decay patterns, when the top-antitop pair decays via $t\bar t\to b W^+\bar b W^-\to b \ell^+\nu_\ell\bar b\ell^-\bar\nu_\ell$.  In order to prove this, we have adopted a 2HDM as theoretical setup, where $X=H_2$ and $H_3$ (or else, $H$ and $A$, respectively, as we have assumed its CP-conserving realization), i.e., spin-0 Higgs bosons with CP-even and CP-odd parity, respectively.  
Upon describing how the individual momenta of the $t\bar t$ system are reconstructed,  in the presence of the additional $X$ decay products, we have adopted a set of kinematic variables giving access to both the overall mass scale of the $X$ system and the mass difference $m_{H_2}-m_{H_3}$ as well as their CP status, with a precision which is competitive with analyses of the full final state $t\bar t H_{2,3}$ (wherein the Higgs bosons are typically extracted by using their $b\bar b$ decays), as we have also proven here.

We thus advocate the prompt deployment of this kind of inclusive studies within the LHC experiments as they are overall less subject to systematics than exclusive ones while offering a similar diagnostic power of the $H_{2,3}$ properties, to such an extent that we have in fact assessed the sensitivity of such approach to both an elementary 2HDM and the C2HDM (a 2HDM stemming from 
Compositeness). In fact, we have been able to show that significant regions of the parameter space of these two BSM scenarios can eventually be accessed. Such a conclusion is  corroborated by a MC  analysis carried out with a level of sophistication comparable to those by ATLAS and CMS and the fact that the described sensitivities can already be achieved by the end of Run 3,  although it will be the statistics afforded by the HL-LHC that will have the ultimate word on the final experimental scope. 

As an outlook, we should mention that work is now in progress in order to assess the validity of our approach in the case of a CPV 2HDM realization, wherein the $H_2$ and $H_3$ states have no definite CP status, rather they are a mixture of the $\pm1$ CPC states, along the lines of Refs.~\cite{Cheung:2020ugr,Azevedo:2022jnd}, including the use of advanced Machine Learning (ML) techniques \cite{Esmail:2024gdc}, again, for the purpose of tensioning elementary \cite{Carena:2001fw} against  Compositeness 
\cite{DeCurtis:2021uqx} versions of CPV 2HDMs. 

\begin{acknowledgments} 
EC is supported by the Funda\c{c}\~ao para a Ci\^encia e a Tecnologia (FCT) under the contract PRT/BD/154189/2022. 
AD is supported by the Carl Trygger Foundation under the project CTS 23:2930.
SM is supported in part through the NExT Institute and the STFC Consolidated Grant No. ST/X000583/1.
{AO acknowledges  funding from FCT through the Strategic Funds projects with references UIDP/04650/2020, UIDB/04650/2020, UID/PRR/04650/2025 and UID/04650/2025.} LDR is supported by the European Union – Next Generation EU
through the research grant number P2022Z4P4B “SOPHYA - Sustainable Optimised PHYsics
Algorithms: fundamental physics to build an advanced society” under the program PRIN 2022
PNRR of the Italian Ministero dell’Universit\`{a} e Ricerca (MUR). LDR and SDC are also supported by the research grant number 20227S3M3B “Bubble Dynamics in Cosmological Phase Transitions” under the program PRIN 2022 of the Italian Ministero dell’Universit\`{a} e Ricerca (MUR).
\end{acknowledgments}

\bibliographystyle{h-physrev}
\bibliography{references}
\nocite{*} 

\raggedbottom 

\begin{widetext}
\appendix
\cleardoublepage

\end{widetext}

\end{document}